\documentclass{aa}  
\usepackage{natbib} 
\bibpunct{(}{)}{;}{a}{}{,}
\usepackage[super]{nth}
\usepackage{graphics}
\usepackage{txfonts}
\usepackage{lscape}
\usepackage{float}
\usepackage[colorlinks,citecolor=blue]{hyperref}
\usepackage{siunitx}
\usepackage{placeins}

\usepackage[authormarkup=none,defaultcolor=darkred]{changes}
\definechangesauthor[color=magenta]{MU}
\hypersetup{
    colorlinks = true,
    linkcolor = blue,
    anchorcolor = blue,
    citecolor = blue,
    filecolor = blue,
    urlcolor = blue
    }

\usepackage{mathrsfs}
\usepackage{multirow,bm}
\usepackage{svg}
\usepackage{ulem}
\usepackage{dirtytalk}

\setcitestyle{notesep={},round,aysep={},yysep={;}}

\newcommand{\unc}[3]{#1$^{+#3}_{-#2}$}

\newcommand{\maui}{\texttt{MAUI}}
\newcommand{\fastwind}{\texttt{FASTWIND}} 
\newcommand{\teff}{$T_{\mathrm{eff}}$}
\newcommand{\logg}{$\log\ g$} 

\begin{document} 

\title{A statistical framework for quantitative spectroscopy of luminous blue stars}

   \author{M. A. Urbaneja\inst{1}}

   \institute{Universit\"at Innsbruck, Institut f\"ur Astro- und Teilchenphysik, Technikerstr. 25/8, 6020 Innsbruck, Austria\\
              \email{miguel.urbaneja-perez@uibk.ac.at}}

   \date{Received ; accepted }
 
\abstract{Quantitative spectroscopy of luminous blue stars relies on detailed non-LTE model 
atmospheres whose increasing physical realism makes direct,
i\-te\-ra\-ti\-ve analyses computationally demanding.}{We introduce \maui~(Machine-learning Assisted Uncertainty Inference), 
a statistical framework designed for efficient Bayesian inference of stellar parameters 
using emulator-based spectral models.}{\maui~employs Gaussian-process-based emulators trained on a limited set of 
non-LTE simulations, combined with Markov Chain Monte Carlo (MCMC) sampling to explore 
posterior distributions. We validate the approach with recovery experiments and demonstrate it 
on Galactic late-type O dwarf and early-type B dwarf/subgiant stars.}{The emulator reproduces the predictions of full atmosphere models within quoted 
uncertainties while reducing computational cost by orders of magnitude. 
Posterior distributions are well calibrated, with conservative coverage across all 
stellar parameters.}{Emulator-driven Bayesian inference retains the accuracy of classical analyses at a
 fraction of the computational expense, enabling posterior sampling that would be prohibitive with direct model 
 evaluations. This positions emulators as a practical tool for high-fidelity spectroscopy of massive stars as atmosphere 
 models grow more demanding.}

   \keywords{Stars: early-type -- stars: fundamental parameters -- stars: atmospheres -- methods: statistical -- methods: numerical -- techniques: spectroscopic }

   \authorrunning{Urbaneja M.A.}
   \titlerunning{Spectroscopic Inference with Emulators and MCMC}   
   \maketitle

\section{Introduction}
 \label{intro-section}

Quantitative spectroscopy of massive stars is central to determining their fundamental parameters, 
surface abundances, and wind properties --- essential ingredients for understanding their structure, 
evolution, and feedback 
\citep[e.g.][]{maeder2000, evans2006, crowther2007, puls2008, langer2012}.
OB-type stars, with extended atmospheres and radiatively driven winds \citep{kudritzki00}, 
require sophisticated non-LTE radiative transfer and line-formation codes to reproduce observed spectra 
(e.g., \citealt{hillier98, pauldrach01, hamann04, puls05}). The computational expense of such models, however, limits their use in iterative fitting procedures and in 
studies requiring dense sampling of highly dimensional parameter spaces.

Optimisation techniques based on metaheuristics
\citep{blumroli2003, talbi2009} have been successfully applied to the spectroscopic analysis of massive stars  
\citep[e.g., genetic algorithm, ][]{mokiem2005}. These methods are attractive for their
robustness in exploring complex parameter spaces without needing
gradients. However, 
while metaheuristics are effective at lo\-ca\-ting global optima, they do not provide a natural means to cha\-rac\-te\-ri\-se the underlying posterior distributions \citep{robert2004montecarlo, posselt2012, schaer2018}. As a result, uncertainty estimates are typically adopted from 
fixed criteria \citep[e.g.,][]{brands2022}, rather than being derived from the full statistical structure of the inference problem. This remains a concern, given the increasing number of free parameters required to model the complex atmospheres of these objects \citep[e.g.,][]{puls2020} and the incomplete understanding of degeneracies within the associated parameter space.

Bayesian inference frameworks offer a rigorous alternative by exploring posterior probability distributions 
conditioned on the data \citep{ford2005, gregory2005}.  Their drawback is the prohibitive number of model evaluations required. Analogous challenges in climate science and cosmology have been addressed with 
statistical emulators --- machine-learning surrogates trained on a limited set of expensive simulations --- 
that deliver fast predictions with calibrated uncertainties 
(e.g., \citealt{heitmann2009, rogers2019, watsonparris2021}).

In this work we present \maui~(Machine-learning Assisted Uncertainty Inference), 
a modular framework that builds a statistical emulator with supervised machine-learning techniques and, 
together with MCMC sampling, enables robust and efficient spectroscopic inference for massive stars. 
While emulators have been used in other areas of astrophysics 
(e.g.\ cosmology, stellar population synthesis, supernova modelling), 
applications to quantitative stellar spectroscopy remain limited. 
\maui~is, to our knowledge, the first framework tailored to the high-dimensional, non-LTE 
parameter spaces of hot stars, integrating the emulator directly into a Bayesian workflow. 
This delivers posterior distributions with full uncertainty quantification at a fraction of the cost 
of traditional analyses. 

This paper is organised as follows. After introducing the methods (Sect.~\ref{sec:methods}), we validate the
framework against direct model evaluations (Sect.~\ref{sec:validation}) and apply it to a set of benchmark 
stars (Sect.~\ref{sec:benchmark}), showing that it recovers stellar and wind pa\-ra\-me\-ters with high fidelity. 
Finally, we conclude discussing future extensions (Sect.~\ref{sec:discussion}). 

\section{Methods}
\label{sec:methods} 

\maui\ is not a replacement for stellar-atmosphere modelling, but a statistical framework that emulates and accelerates the in\-fe\-ren\-ce process while fully preserving the physical content of the underlying models. The framework combines synthetic spectra computed with an atmosphere code with statistical components for dimensionality reduction and Gaussian-process emulation, together with Bayesian inference for parameter estimation and uncertainty quantification. The following subsections describe the construction of the training grid, the compression of the stellar spectra and emulator training, and the likelihood formulation used in the Bayesian analysis.

\subsection{Model atmosphere and spectral synthesis}
For the applications presented in this work, we employ the non-LTE model atmosphere and spectral 
synthesis code \fastwind~(version 10), originally introduced by \cite{santolaya-rey1997}, and subsequently extended by 
\cite{puls05}, and \cite{rivero-gonzalez11}. A detailed account of the current version of the code, including  
comparisons with alternative codes, is provided by \cite{carneiro16}. \fastwind~solves the radiative transfer problem for spherically symmetric, expanding stellar 
atmospheres, under the assumptions of stationary outflow and chemical homogeneity. The code simultaneously ensures statistical 
equilibrium and energy conservation, treating line-blocking and line-blanketing effects
in a consistent, well-tested approximate way, thus reducing the computational effort by more than a factor of 10 to 20 
compared to an exact treatment.
The density structure is computed by assuming hydrostatic equilibrium in the deep photosphere and applying the equation of 
continuity in the wind domain, where the velocity field follows a standard $\beta$-type law. A continuous transition between the 
quasi-hydrostatic photosphere and the accelerating wind is imposed.

\fastwind~allows for parameterised treatments of wind inhomogeneities 
and X-ray emission from embedded shocks. In the standard microclumping approach, 
clumps are assumed optically thin, while macroclumping relaxes this assumption 
and reduces effective opacities in lines and continua 
\citep[e.g.][]{hillier1991, hamann1998, puls06, oskinova2007, sundqvist2010}. 
We neglect any kind of wind inhomogeneities: optical spectra provide li\-mi\-ted 
leverage on clumping (H$\alpha$ -- and HeII 4686 in O-stars -- being the only 
sensitive lines), and our results should thus be interpreted as smooth-wind 
models, with the usual systematic uncertainties in absolute mass-loss rates. 
Similarly, while \fastwind~can account for X-ray/EUV emission from 
wind-embedded shocks \citep{carneiro16, puls2020}, such effects are negligible 
for the photospheric lines analysed in this work.

Microturbulence enters the modelling at two distinct stages: first, in the 
construction of the stellar atmosphere (stratification and level populations), 
and second, in the computation of the formal solution (emergent spectrum). 
For the atmosphere calculations we adopt a fixed value of 
$10\,\mathrm{km\,s^{-1}}$ for the additional broadening of the line-profiles, though 
neglecting any turbulent pressure in the hydrostatic/-dynamic
description. 
In the formal solution $\xi$ is treated as a free 
parameter and allowed to vary from model to model. Although \fastwind\ also offers 
a depth-dependent prescription for $\xi$ in the formal solution, 
its impact is limited mainly to UV resonance lines formed in the wind and is 
negligible for the optical photospheric diagnostics considered here. We therefore 
assume a depth-independent value of $\xi$ in the calculation of the emergent profiles.

Consequently, each \fastwind~simulation is specified by a set of input parameters: 
effective temperature \teff, effective surface gravity \logg, and stellar radius $R_\star$ 
(all evaluated at $\tau_{\rm Ross}=2/3$); microturbulent velocity $\xi$; exponent $\beta$ 
of the adopted velocity law; mass-loss rate $\dot{M}$; terminal wind speed $v_\infty$; 
and a set of elemental abundances.

\subsection{Model atoms}
Atomic data play a central role in non-LTE stellar atmosphere modelling. Modern codes provide a sophisticated numerical framework to solve the radiative transfer and statistical e\-qui\-li\-brium equations, but the accuracy of the resulting atmospheric structures and synthetic spectra ultimately depends on the qua\-li\-ty and completeness of the underlying atomic data \citep{przybilla2010,hillier2011}. In this sense, non-LTE codes are largely data-driven: the code itself implements the physics and nu\-me\-ri\-cal methods, but the computed populations, line strengths, and emergent spectra are only as reliable as the atomic input. Consequently, differences in oscillator strengths, cross-sections, or e\-ner\-gy levels can propagate into systematic differences in inferred stellar parameters and abundances.

To achieve high computational efficiency, \fastwind~distinguishes between two classes of elements: so-called {\it explicit} and {\it background} elements \citep{puls05}. Background elements, which are important for the overall atmospheric structure but not used as direct diagnostics, are treated approximately using a fixed atomic database. For the strongest transitions of background elements between carbon and zinc, the radiative transfer is solved in the comoving frame, while the remaining metal lines are treated either by a conventional static radiative transfer in
photospheric regions or in the Sobolev approximation. This database (based on \citealt[][]{pauldrach1998}) is an integral part of the code distribution and remains unchanged across different applications of \fastwind.

In contrast, explicit elements—those whose lines are employed as diagnostics in quantitative spectroscopy—are des\-cri\-bed using detailed, user-supplied atomic data files. These elements are treated with higher precision, including co-moving frame radiative transfer for all relevant transitions. This distinction ensures that spectral features used in fitting procedures are modelled as accurately as possible while retaining overall computational efficiency. 
Table~\ref{tab:atoms} summarises the model atoms adopted in this study, including the 
original references where they were first described. The atomic data for H, He, and N~{\sc ii/iii/iv} are 
identical to those used in previous \fastwind~stu\-dies 
\citep{rivero-gonzalez12,carneiro2019}, ensuring direct continuity with earlier ana\-lyses. 
In contrast, the models for N~{\sc i}, C\,\textsc{ii}, Mg\,\textsc{i/ii}, and Si\,\textsc{ii--iv} 
are based on newly implemented 
datasets (see Table~\ref{tab:atoms}), while for the O\,\textsc{i--iii}
model we use the same data as in our earlier work 
\citep{urbaneja03,urbaneja05b,urbaneja11,urbaneja17}. 
The C\,\textsc{iii--iv} model atoms have been updated in the meantime (J.~Puls, priv.~comm.).

\begin{table}[h!]
\centering
\caption{Model atoms used in the non-LTE calculations.\label{tab:atoms}}
\begin{tabular}{lccc}
\hline
Ion & Terms & Transitions & Reference \\
\hline
H                       & 20        & 380   & [1] \\
He~{\sc i/ii}       & 49/20   & 2048 & [1] \\
C~{\sc ii/iii/iv}    & 67/70/50        & 6194           & [2,3] \\
N~{\sc i/ii/iii/iv}   & 89/50/41/50  & 4769           & [4,5] \\
O~{\sc i/ii/iii}      & 51/52/61        & 3531           & [6,7,8] \\
Mg~{\sc i/ii}       &  88/37            & 5189           & [9] \\
Si~{\sc ii/iii/iv}    & 39/69/35        & 4946           & [10] \\
\hline
\end{tabular}
\tablebib{
(1)~\citet[][]{jokuthy2002}; 
(2)~\citet[][, C~{\sc ii}]{nieva2008}; 
(3)~\citet[][, C~{\sc iii/iv}]{carneiro18}; 
(4)~\citet[][, N~{\sc i}]{przybilla2001}
(5)~\citet[][, N~{\sc ii/iii/iv}]{rivero-gonzalez12}; 
(6)~\citet[][, O~{\sc i}]{przybilla2000};
(7)~\citet[][, O~{\sc ii}]{becker1988}; 
(8)~\citet[][, O~{\sc iii}]{urbaneja2004} 
(9)~\citet{przybilla2010}; (10)~Przybilla \& Butler (in prep.).
}
\tablefoot{The number of transitions corresponds to the total number of
bound-bound transitions (radiative plus collisional line transitions).}
\end{table}

\subsection{Definition of the training simulations}
\label{sec:grid-design}

The construction of the training grid is a crucial step, since the extent and sampling of the 
physical parameter space directly control how accurately the emulator can reproduce model spectra 
and how broadly it can be applied to real observations.
Three main considerations guide the adopted parameter ranges:  

\begin{enumerate}
    \item {Spectral sensitivity.}
     
     The {\it effective parameters} explored by the grids are set such that the associated spectral 
     diagnostics remain sensitive across the grid. Because optical wind diagnostics are pri\-ma\-ri\-ly sensitive to 
     the combination of mass-loss rate, stellar radius, and terminal velocity through the optical-depth invariant
     $Q = \dot{M} / (R_\star\,v_\infty)^{1.5}$ \citep{puls1996}, there is no practical gain in treating 
     $R_\star$ and $v_\infty$ as independent free parameters\footnote{This statement applies to the parameter space covered in the present work.}. 
     Instead, the stellar radius is assigned consis\-tently from the extended flux-weighted 
     gravity–luminosity relation (FGLR) of \citet{kudritzki2020}, which links the flux-weighted gravity 
     $\log g_{\rm F} = \log g - 4\log(T_\mathrm{eff}/10^4\,\mathrm{K})$ of an object to its luminosity. The terminal 
     velocity is then specified through a scaling with the escape velocity, 
     $v_\infty = f\left(T_\mathrm{eff}\right)\,v_{\rm esc}$, adopting the empirical factor $f\left(T_\mathrm{eff}\right)$ from \citet{kudritzki00}. 
     This prescription ensures physically motivated values of $R_\star$ and $v_\infty$ while keeping the inference effectively 
     focused on the wind-invariant
     parameter $Q$, which instead of $\dot{M}$ and $v_\infty$ enters the definition of our grids. We note that the actual $\dot{M}$-value that 
     enters the model atmosphere input is calculated from the above equation relating $\dot{M}$ with $Q$, $R_\star$ and $v_\infty$. 
     
We note that the extended FGLR of \citet{kudritzki2020} is an empirical relation based, for the most part, on detached eclipsing binaries (see the reference for details) and is therefore not tied to a specific set of stellar evolution mo\-dels. While the calibration sample discussed in the aforementioned work focuses on masses up to $\sim$20~$M_\odot$, the pa\-ra\-me\-ter range relevant for the present study is well bracketed by the systems included in the underlying eclipsing-binary catalogue. Moreover, given the weak sensitivity of optical wind diagnostics to the stellar radius when working with normalised spectra, any residual uncertainty associated with the adopted radius assignment has only a minor impact on the inferred wind-invariant parameter $Q$.
    
    \item {Astrophysical relevance.}  
    
The ranges are motivated by the typical properties of the target stars under investigation: they were designed to fully cover the stellar and wind properties expected for the stars analysed in this work. 
For example, the grid spans microturbulent velocities from \(\xi = 0\) to \(20~\mathrm{km\,s^{-1}}\), encompassing the range typically 
inferred from metal lines in Galactic OB dwarfs and giants \citep[e.g.,][]{przybilla2008,simondiaz2010}. Similarly, the wind parameters 
($\beta$ and $\log Q$) were varied across ranges sufficient to reproduce the H$\alpha$ line morphologies observed in these stars 
\citep[e.g.,][]{holgado18,carneiro2019}.

     \item {Computational feasibility.}
  
Rather than employing a fine regular mesh, we use a space-filling sampling strategy to efficiently explore the multidimensional parameter space. 
Such designs provide good co\-ve\-ra\-ge of each parameter range while avoiding redundancy between sampling points.

\end{enumerate}

The training grid consists of 995 \fastwind~simulations, 
each corresponding to a unique combination of stellar and wind parameters 
drawn from a maximum-projection Latin-hypercube design \citep{mckay1979, stein1987, joseph2015}, which ensures 
uniform coverage of the parameter space without requiring a regular grid. 

The stellar parameters covered are the effective temperature \teff,
surface gravity \logg, 
microturbulent velocity $\xi$ (formal solution), wind-strength parameter $\log Q$, velocity law 
exponent $\beta$, and individual abundances of He, C, N, O, Mg, and Si. 
The adopted parameter ranges (Table~\ref{tab:grid_ranges}) cover the observational domain of O9–B3 
dwarfs and giants, but are extended in some cases beyond typical values. 
For example, the helium abundance is allowed to reach \(\mathrm{He/H}=0.07\), 
slightly below the primordial value, not because such low abundances are expected, 
but to ensure that the emulator samples a 
sufficiently wide domain of the physical model behaviour. By exposing the emulator to 
the full range of spectral responses, including regions only marginally 
relevant for our targets, the statistical model can better learn the underlying 
dependencies on the physical parameters. This strategy reduces edge effects and 
leads to more robust interpolations within the astrophysically relevant 
parameter space, which is where the inference is ultimately performed.

The number of training models was chosen based on the expected smoothness of the spectral 
response and the di\-men\-siona\-li\-ty of the parameter space. 
The adopted parameter ranges therefore define the applicability domain of the present emulator: 
its reliability is ensured only within the boundaries of the training grid, and 
predictions beyond these limits should be regarded as extrapolations and treated with caution.

\begin{table}
\caption{Parameter ranges adopted for the training grid of Galactic late O/early B-type dwarfs/giants.}
\label{tab:grid_ranges}
\centering
\begin{tabular}{lcc}
\hline\hline
Parameter & Minimum & Maximum \\
\hline
\teff\ [K]                    & 17000  & 38000 \\
$\log g$ [dex]           & 3.25  & 4.40 \\
He/H (number)        & 0.07  & 0.20\\
$\xi$  [km\,s$^{-1}$] & 0 & 20 \\
$\beta$                    & 0.8  & 2.0 \\
$\log Q$            & -15.00  & -12.80 \\
C                   & 7.70  & 8.70 \\
N                   & 7.00  & 8.70 \\
O                   & 8.26  & 9.26 \\
Mg                  & 7.06  & 8.06 \\
Si                  & 7.00  & 8.00 \\
\hline
\end{tabular}
\end{table}

\subsection{Dimensionality reduction}
\label{klt-section}

A core challenge in emulating stellar spectra lies in their inherently high 
dimensionality: each synthetic spectrum typically consists of thousands of 
flux values sampled over a finely spaced wavelength grid. This high dimensionality 
poses significant computational and statistical challenges for any supervised 
learning technique, as models must capture complex 
spectral variations while avoiding overfitting and maintaining efficiency. In 
particular, regression methods must cope with increased data sparsity and longer 
training times as the output dimensionality grows. These challenges are especially 
pronounced in surrogate modelling applications where the 
goal is to replace computationally expensive simulations with fast emulators 
trained on a li\-mi\-ted number of examples \citep[][]{bengio2013representation,grover2018learning}.

To alleviate this problem, we construct an embedding-- a lower-dimensional representation of the synthetic spectra -- using Principal Component Analysis \citep[PCA,][]{connolly1995}, whose mathematical foundations are rooted in the Karhunen–Lo\`eve Transform \citep[KLT,][]{loeve1978}. Very briefly, the KLT identifies an orthonormal basis that diagonalises the covariance 
matrix (see below), with eigenvectors capturing the dominant modes of spectral variation. Projecting each spectrum onto this basis yields a set of coefficients, which, along with the eigenvectors, constitutes the lower-dimensional representation of the original space.\\

In practice, before calculating the embedding, we preprocess the synthetic spectra by 
interpolating them onto a common wavelength grid. 
The mean\footnote{Arithmetic mean.} spectrum, \(\bar{\mathbf{x}}\), 
is subtracted from all flux vectors to centre the data. 
The resulting centred data matrix \(\mathbf{X} \in \mathbb{R}^{N \times D}\), 
where \(N\) is the number of spectra (i.e.\ models) and \(D\) the number of wavelength points, 
has a sample covariance matrix \(\mathbf{C} \in \mathbb{R}^{D \times D}\) defined as
\[
\mathbf{C} = \frac{1}{N - 1}\,\mathbf{X}^\mathsf{T}\mathbf{X}.
\]

The KLT basis vectors \(\mathbf{u}_k\) are the eigenvectors of 
\(\mathbf{C}\), while the associated eigenvalues \(\lambda_k\) measure the variance of the training 
spectra along the corresponding eigenvector directions. In other words, each \(\lambda_k\) quantifies the amount of variance 
explained by the corresponding principal component, with larger eigenvalues indicating 
components that capture more of the structure in the input data. 

The matrix of eigenvectors is denoted 
\(\mathbf{U} = [\,\mathbf{u}_1, \ldots, \mathbf{u}_D\,] \in \mathbb{R}^{D \times D}\),
and the corresponding projections
\[
\mathbf{Z} = \mathbf{X}\mathbf{U}, \qquad \mathbf{Z} \in \mathbb{R}^{N \times D},
\]
yield the coefficients \(z_{ik}\).  

Each original spectrum \(\mathbf{x}_i\) can be reconstructed as
\[
\mathbf{x}_i = \bar{\mathbf{x}} + \sum_{k=1}^{D} z_{ik}\,\mathbf{u}_k ,
\]
where \(z_{ik}\) is the coefficient of the \(k\)-th component for spectrum \(i\).

By retaining only the first \(K\) components—those associated with the largest eigenvalues,
chosen to explain a given fraction of the total variance—the dimensionality of the output space 
is reduced from \(D\) to \(K\), where \(K \ll D\). 
The simulated spectra in the training dataset can then be approximated as
\[
\hat{\mathbf{x}}_i = \bar{\mathbf{x}} + \sum_{k=1}^{K} z_{ik}\,\mathbf{u}_k,
\]
where \(\hat{\mathbf{x}}_i\) denotes the reconstructed approximation of the original spectrum 
\(\mathbf{x}_i\).
This reconstruction is approximate, as the truncated sum omits the higher-order components; 
the resulting error corresponds to the residual variance not captured by the first \(K\) principal components.

While nonlinear dimensionality reduction methods are increasingly popular in machine 
learning, we adopt a linear approach for its interpretability. The linear transform ensures 
that reconstructed spectra remain physically meaningful and avoids spurious artefacts that 
may arise in nonlinear mappings. Still, linear embeddings, such as those used here, may 
miss subtle nonlinear dependencies in certain regions of parameter space. Methods like 
autoencoders or manifold learning \citep[e.g., ][]{hinton2006reducing, cunningham2015}
could, in principle, capture such curved structures more 
faithfully, though at the cost of reduced interpretability and increased sensitivity to 
training and hyperparameter choices \citep[][]{saxe2019information}. In this work, we prioritise 
physical transparency and stability, while leaving nonlinear or hybrid extensions to future 
developments.

\begin{figure*}[h]
  \centering
  \includegraphics[width=0.9\textwidth]{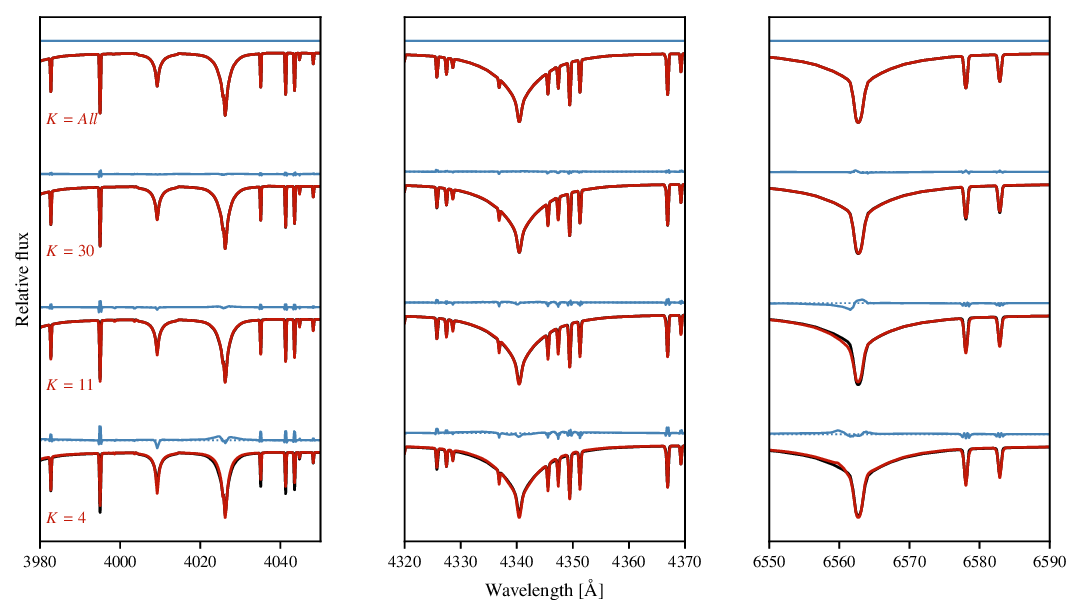}    
\caption{Reconstruction of a representative training spectrum for different truncation levels. 
Each panel corresponds to a wavelength range and displays the original spectrum (black) together with 
reconstructions obtained using $K=4,\,11,\,30$, and all principal components (red), vertically offset for clarity. 
The blue curves show the corresponding residuals (reconstruction$-$original) plotted above each case. 
The examples illustrate the progressive improvement in the reconstructed line profiles with increasing $K$.
The selected windows contain both strong and weak diagnostic features (H, He, and metals), 
including the wind–sensitive H$\alpha$ region.}

  \label{fig:klt_truncation}
\end{figure*}

\begin{figure*}[h]
  \centering
  \includegraphics[width=0.9\textwidth]{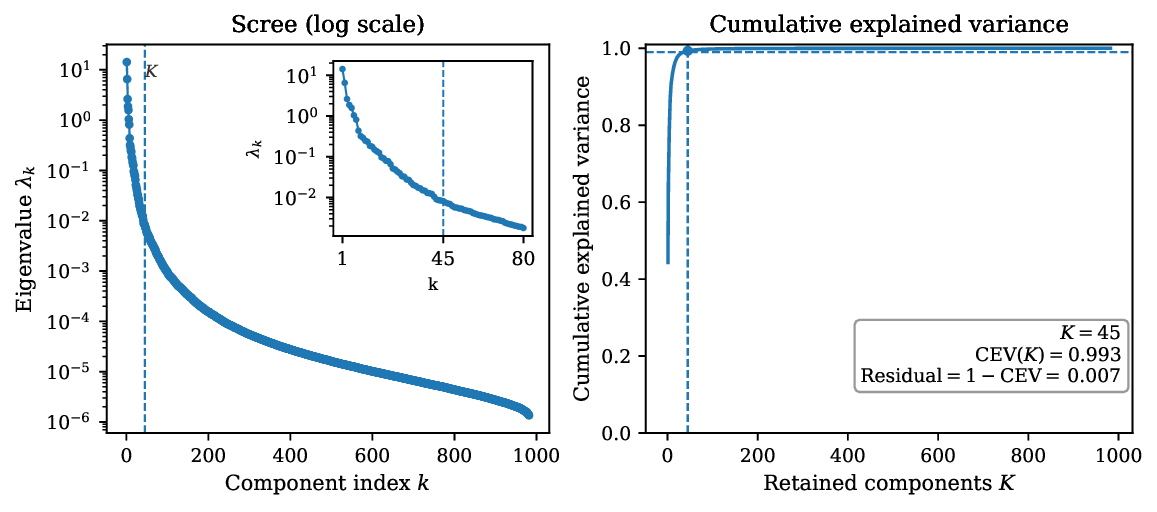}  
\caption{Principal-component analysis diagnostics. 
Left: Scree plot of eigenvalues~$\lambda_k$ (log scale); 
the vertical dashed line marks the adopted truncation at $K=45$, 
beyond which the eigenvalues decay rapidly (grey-shaded region, inset zoom). 
Right: Cumulative explained variance, showing that the first 
$45$ components capture over $99\%$ of the total spectral variance.}
  \label{fig:pca_panel}
\end{figure*}

To assess the fidelity of the compact representation independently of the emulator, we use the
full training set to compute the reconstruction root-mean-square error (RMSE). Each spectrum in
the training grid is projected onto the first \(K\) basis vectors to obtain the corresponding
coefficients \(z_{ik}\), reconstructed, and compared to the original simulation. The RMSE is defined as
\[
\varepsilon_{\mathrm{KLT}}
= \sqrt{\frac{1}{N}\sum_{i=1}^{N}
  \frac{\bigl\|\mathbf{f}_i -
        \sum_{k=1}^{K} z_{ik}\mathbf{u}_k\bigr\|^2}
       {\|\mathbf{f}_i\|^2}},
\]
where \(\mathbf{f}_i\) is the original \(i\)-th spectrum.

The practical impact of this dimensionality reduction is illustrated in Fig.~\ref{fig:klt_truncation},
which shows how well spectra can be reconstructed from truncated sets of components. Four
cases—\(K=4,11,30\), and \(995\) (all)—are displayed.  A total of \(K=45\) components are
required to keep the global RMSE below \(3\times10^{-3}\); these account for 99.3\,\% of the total
variance in the training set. This threshold is not arbitrary: reconstructions at this level faithfully
reproduce the spectral regions most relevant for diagnostics while keeping the dimensionality low
enough for efficient regression. Tests with fewer components led to noticeable degradation in line
cores, whereas including more yielded negligible gains at significant computational cost.

The statistical justification for this truncation is summarised in
Fig.~\ref{fig:pca_panel}. The left-hand panel shows the eigenvalue spectrum, which exhibits a
clear knee near \(K\simeq45\), beyond which additional components contribute negligibly to the
variance. The right-hand panel displays the cumulative explained variance, confirming that the
first 45 components capture more than 99\,\% of the variance in the training set. This validates the
choice of \(K\) adopted for all subsequent analyses.

\begin{figure*}[h]
  \sidecaption
  \includegraphics[width=12cm]{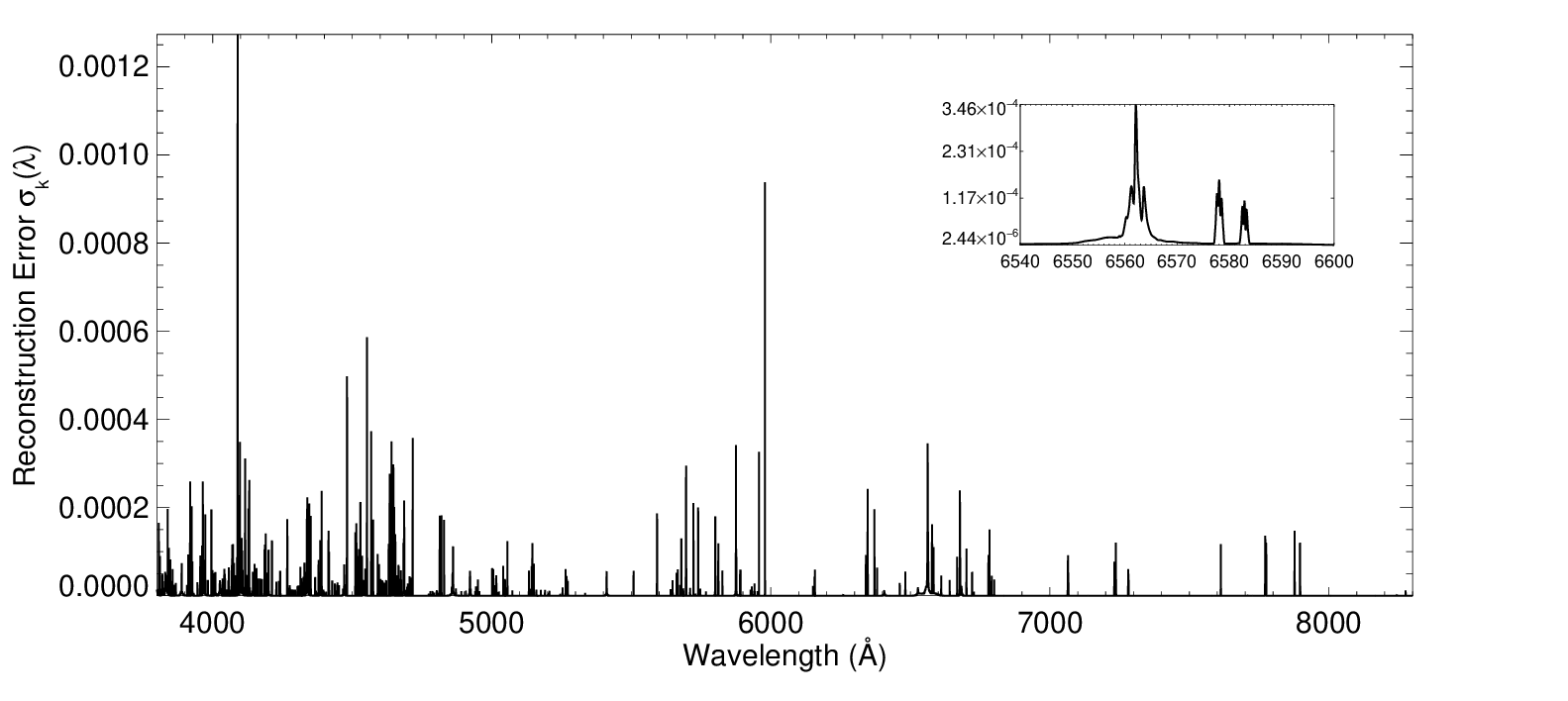}
\caption{Wavelength-dependent reconstruction error, \(\sigma_K(\lambda)\), for the case \(K=45\). The inset zooms on the region around 
H$\alpha$. Interestingly, the largest deviation occurs at the Si\,\textsc{iv}/O\,\textsc{ii} blend near 4089\,\AA, rather than at H$\alpha$. 
}
  \label{fig:klt_truncation_wav}
\end{figure*}

The wavelength-dependent reconstruction error for \(K=45\) is shown in Fig.~\ref{fig:klt_truncation_wav}. 
This error, \(\sigma_\mathrm{T}(\lambda)\), forms part of the total error budget considered during the inference process 
(see Sect.~\ref{sec:mcmc}). Interestingly, and somewhat counterintuitively, the largest deviation does not 
occur at H$\alpha$, but at the Si\,\textsc{iv}/O\,\textsc{ii} blend near 4089\,\AA. 
Because of its relatively large reconstruction error, this feature is assigned a lower weight 
in the likelihood evaluation.

\subsection{Learning the PC coefficients with Gaussian processes}
\label{regression-section}

After compressing each synthetic spectrum into $K$ principal--component (PC) coefficients $\{z_k\}_{k=1}^K$, the remaining task is functional interpolation: for any stellar--parameter vector $\boldsymbol{\theta}$, we wish to predict the corresponding set of coefficients $z_k(\boldsymbol{\theta})$ that would be obtained at that location in parameter space.

For each retained component $k$, we therefore infer a smooth mapping
\[
\boldsymbol{\theta} \rightarrow z_k(\boldsymbol{\theta}),
\]
using the spectra in the training set. Once these $K$ mappings are known, the coefficients $\{z_k(\boldsymbol{\theta}_*)\}$ for a new parameter vector $\boldsymbol{\theta}_*$ can be predicted, and the corresponding emulated spectrum reconstructed by combining the eigenvectors with these coefficients, exactly as for any original simulation in the grid.

For a given PC, there exists a family of possible functions that could
reproduce the corresponding coefficint values from the training grid. Rather than 
seeking an explicit functional form, our goal is to predict---with quantified uncertainty---the most probable coefficient value for a new parameter 
vector $\boldsymbol{\theta}_*$. Gaussian--process (GP) regression \citep{RasmussenWilliams2006} provides a natural framework for this 
task, as it defines a probability distribution over functions consistent with the data.
Each PC coefficient is modelled independently, which is justified by the orthogonality of the PC basis: the variables are uncorrelated by construction, so modelling them independently preserves information. Although finite grid sampling, truncation to $K$ components, and mild nonlinearities can introduce small residual correlations, these effects are empirically negligible for reconstruction and inference: the
 off-diagonal elements of the covariance matrix are typically below 2-3 \% of the total variance, confirming that the assumption of independence is well justified.

The GP prior assumes that the underlying function is smooth and continuous, with continuous derivatives. It is fully specified by a mean function 
and a covariance kernel, the latter describing how the similarity between two predicted coefficients depends on the distance between their 
corresponding models in parameter space. The exact form of the covariance kernel and its hyperparameters---typically an amplitude and a set of 
characteristic length scales (one per model parameter)---determine how rapidly the coefficients are allowed to vary across the grid. The kernel 
should therefore reflect the expected smoothness of the underlying physical behaviour.

We tested various stationary kernels---whose covariance depends only on separation in parameter space---including squared--exponential, Matérn, and rational--quadratic forms, as well as their linear combinations. The best performance was achieved with a squared--exponential term (capturing smooth global trends) plus a white--noise term (absorbing numerical noise), ensuring accurate, stable interpolation without overfitting. In all cases, the noise amplitude is orders of magnitude smaller than the main covariance term.

To account for differing sensitivities of the spectrum to each input parameter, we adopt an automatic--relevance--determination (ARD) kernel 
\citep[see, e.g.,][]{RasmussenWilliams2006} with a separate length scale per parameter. The hyperparameters are learned from the training data by maximising the marginal likelihood. Before training, both the target coefficients and the input parameters are standardised to zero mean and unit variance to improve numerical conditioning.

The predictive variance of each GP, $\sigma_k^2(\boldsymbol{\theta})$, propagates into the emulator uncertainty at each wavelength as
\begin{equation}
\sigma_{\mathrm{emu}}^2(\lambda)
= \sum_{k=1}^K \sigma_k^2(\boldsymbol{\theta})\,u_k^2(\lambda),
\end{equation}
where $\mathbf{u}_k(\lambda)$ is the corresponding PC eigenvector. This expression follows directly from 
the linear reconstruction of the e\-mu\-la\-ted spectrum, assuming the PC coefficients are modelled as 
independent. The emulator variance contributes directly to the total error budget used in the model--likelihood 
evaluation (see Sect.~2.6). Unlike the truncation error introduced earlier or the observational noise, this term is model--dependent, i.e. it varies with $\boldsymbol{\theta}$.

Outside the boundaries of the training grid, the GP predictive variances $\sigma_k^2(\boldsymbol{\theta})$ 
increase rapidly, leading to a correspondingly large emulator variance $\sigma_{\mathrm{emu}}^2(\lambda)$. This 
behaviour provides a na\-tu\-ral warning against extrapolation and is one of the main reasons we restrict the emulator to the parameter domain covered by the training set.

\subsection{Parameter inference}
\label{sec:mcmc}

The ultimate goal of quantitative spectroscopy is to determine the atmospheric parameters and surface chemical abundances that characterise an observed spectrum by comparing it with synthetic ones. This process relies on three key ingredients: (1) the observed data, (2) a collection of synthetic spectra, and (3) the metric that quantifies how well a model reproduces the observations. Each of these components directly influences the outcome of the inference. On the observational side, both the quality of the data (e.g., signal-to-noise ratio) and the wavelength co\-ve\-ra\-ge determine the available constraints. On the modelling side, the fidelity of the underlying physics and the chosen parameterisation shape the predictive power of the synthetic spectra. Finally, the definition of the distance metric---such as the choice of diagnostic lines and/or the use of weighting schemes---plays a critical role in guiding the inference.   
A full discussion of these aspects is beyond the scope of the present work, 
but fo\-re\-sha\-do\-wing the results in Sect.~\ref{sec:results}, we stress that they 
all contribute to the robustness of the derived parameters. As will become 
apparent in the comparison with literature values, some differences may arise 
from the characteristics of the observational material or the specific models 
employed, in addition to methodological choices.

The inference task is naturally formulated within a Bayesian framework, where the aim is to
determine the full posterior pro\-ba\-bi\-li\-ty distribution of the
parameters, $\boldsymbol{\theta}$, given the data, $D$. 
Bayes’ theorem provides the formal connection (see, e.g., \citealt{gregory2005} for an 
introduction to Bayesian methods in as\-tro\-no\-my):
\begin{equation}
    p(\boldsymbol{\theta} \,|\, D) = \frac{\mathcal{L}(D\,|\,\boldsymbol{\theta})\,\pi(\boldsymbol{\theta})}{Z},
\end{equation}
where $\mathcal{L}(D \,|\, \theta)$ is the likelihood function, quantifying how well a given parameter 
set reproduces the observed spectrum, $\pi(\theta)$ denotes the prior distribution, and 
\begin{equation}
    Z = \int \mathcal{L}(D \,|\,\boldsymbol{\theta})\,\pi(\boldsymbol{\theta})\,d\boldsymbol{\theta}
\end{equation}
is the Bayesian evidence (normalisation constant). The evidence ensures that the posterior 
distribution integrates to unity, but since we are not comparing models with 
different dimensionalities or physical assumptions, $Z$ plays no role in the present 
analysis. The inference therefore depends only on the product of likelihood and prior.

The posterior distribution is analytically intractable, and we therefore rely on Markov Chain Monte Carlo (MCMC) sampling to obtain representative draws from it. Specifically, we employ a Metropolis--Hastings algorithm \citep{Metropolis1953,Hastings1970}, which constructs a Markov chain whose stationary distribution converges to the target posterior. This approach allows us to efficiently explore the probability landscape, quantify uncertainties, and diagnose correlations or degeneracies between parameters \citep{gregory2005}.  

The likelihood is evaluated by comparing the observed spectrum $F^{\rm obs}$ to the theoretical spectrum 
(in this case, emulated) $\hat{M}(\boldsymbol{\theta})$ for a trial parameter vector. It takes the form

\begin{equation}
- \ln \mathcal{L}(\mathcal{D} \,|\, \boldsymbol{\theta}) = 
\frac{1}{2} \sum_{\lambda} 
\left[ \frac{F^{\mathrm{obs}}_{\lambda} - \hat{M}_{\lambda}(\boldsymbol{\theta})}{\sigma_{\lambda}} \right]^{2} 
+ \ln \!\left( 2 \pi \sigma_{\lambda}^{2} \right).
\end{equation}

with the total uncertainty per wavelength point given by 
\(\sigma_\lambda^2 = \sigma_{\rm obs}^2 + \sigma_{\rm emu}^2 + \sigma_{\rm T}^2\). 
Here, \(\sigma_{\rm obs}\) represents the observational noise (e.g.\ photon noise), 
\(\sigma_{\rm emu}^2\) is the predictive variance of the emulator, 
and \(\sigma_{\rm T}\) accounts for the residual wavelength-dependent reconstruction error 
introduced by truncation of the PCA basis. 
This formulation assumes that the forward model is unbiased, 
i.e.\ that systematic offsets between the simulations and the observations are 
negligible compared to the stochastic error terms. 
In practice, such systematics may arise from limitations in the underlying model physics or input data; 
they are not explicitly modelled here but could, in principle, be represented 
through an additional \textquotedblleft model discrepancy\textquotedblright\ term $\sigma_{\mathrm{md}}$ 
in the likelihood. Future extensions of the framework will incorporate such effects.

Because the emulator is computationally inexpensive to evaluate, it enables extensive MCMC exploration without invoking full radiative transfer calculations at each step. Priors are taken to be uniform within physically motivated ranges, consistent with the limits of the training grid, and we explicitly forbid extrapolation beyond the domain where the emulator is valid.  Posterior distributions were sampled using single MCMC chains of $5\times10^{4}$ steps, discarding the first half as burn-in. Convergence was assessed using trace plots to verify stationarity and mixing, and by computing the effective sample size (ESS) of each parameter to quantify sampling efficiency. This setup was found to yield stable posterior estimates for all inferred parameters.

This inference setup is applied consistently throughout this work, first to the 
validation simulations (Sect.~\ref{sec:validation}) and then to the analysis of 
observed spectra (Sect.~\ref{sec:results}).

\subsection{Computational requirements}
\label{sec:comp-req}

All components of the emulator and inference pipeline are implemented in \textsc{IDL}\,8.8.0 (single-threaded) on Rocky Linux~9.4 (Blue Onyx);
we rely on \textsc{IDL}'s native linear-algebra.

We denote by $N_{\mathrm{mod}}$ the number of models in the training grid, by $N_{\mathrm{pix}}$ the wavelength pixels effectively used per spectrum (after masking/rebinning), by $N_{\mathrm{pc}}$ the number of retained principal components, and by $N_{x}$ the number of diagnostic windows used at inference. In practice, increasing $N_{\mathrm{mod}}$ chiefly raises training time; larger $N_{\mathrm{pix}}$ mainly affects the initial compression and the cost to reconstruct spectra; and the per-evaluation runtime scales roughly with $N_{x}$ and the pixels per window because of the instrumental, rotational, and macroturbulent convolutions. Adding more atmospheric parameters typically necessitates a denser grid (larger $N_{\mathrm{mod}}$) to maintain coverage and often introduces additional spectral variance, which in turn may require a larger $N_{\mathrm{pc}}$ to preserve reconstruction fidelity.

During inference, each likelihood evaluation comprises emulator prediction and spectral reconstruction, followed by convolution within $N_{x}$ diagnostic windows ($N_{x}\!\ge\!60$ here) with the instrumental line–spread function plus rotational and macroturbulent kernels, resampling to the observed pixels, masking, and the likelihood calculation. The measured mean end-to-end latency per evaluation is $t_{\rm eval}\!\approx\!0.8$\,s.

The one-off cost of generating the training grid can be estimated as
follows. A single \fastwind\ model takes on average $\sim$1.0\,h of CPU time (with mild variation across parameter space). A $\sim$1000-model grid therefore represents $\sim$1000\,CPU-hours; wall-clock is reduced by parallel runs on multi-core nodes. This cost is paid once and amortised across all targets.

For the grid used here ($N_{\mathrm{mod}}=995$,
$N_{\mathrm{pix}}=\,90000$, $N_{\mathrm{pc}}=45$, $N_{x}\!\ge\!60$), emulator training took $\sim$9.5\,h with a peak memory footprint of $4.5\,\mathrm{GB}$ on a dual-socket AMD EPYC\,7302 workstation (64 logical CPUs, $\sim$1.0\,TiB RAM). With $t_{\rm eval}\!\approx\!0.8$\,s, a run with $5\times10^{4}$ evaluations requires $\sim$11--12\,h of wall-clock on a single thread (linear scaling with evaluation count). For com\-pa\-ri\-son, the same number of direct model evaluations would amount to $\sim5\times10^{4}$\,CPU-hours—computationally intensive for MCMC-scale sampling—even before accounting for the one-off training of the emulator.\footnote{Forward model simulations can be parallelised across cores, reducing wall-clock time (e.g. on our 32--64-way machine, $5\times10^{4}$ runs would still require $\sim$32--65\,days). The emulator builds on \fastwind~to enable MCMC-scale posterior sampling; it is intended as a statistical accelerator rather than a replacement for the underlying physical modelling.}.

\section{Validation of the emulator}
\label{sec:validation}

The purpose of this section is to demonstrate that replacing direct forward simulations with the
emulator does not compromise the inference of stellar parameters and abundances. Put differently,
we aim to verify that the use of the emulator is effectively e\-qui\-va\-lent to analysing spectra with the
original atmosphere code. To this end, we carried out a series of controlled recovery experiments:
synthetic spectra were generated at known parameter values, treated as mock observations, and
then analysed with the emulator-based MCMC framework. Comparing the re\-co\-ve\-red posterior
distributions with the true input values allows us to identify potential biases, quantify the reliability
of the quoted uncertainties, and assess whether the emulator reproduces the inference process
faithfully across the full parameter space.

\subsection{Setup}

The validation is based on an independent set of 400 \fastwind~simulations that were not used
for training. These test points were selected via an independent Latin hypercube design, ensuring
broad coverage of the parameter space. The adopted number of simulations represents a compromise
between statistical robustness and computational feasibility. Several hundred points are sufficient
to obtain stable estimates of reconstruction errors, biases, and coverage fractions across the
11-dimensional parameter space (e.g., \citealt{mckay1979, stein1987}), while avoiding the high cost
of much larger validation sets. Smaller test sets ($N \lesssim 100$) proved noticeably noisier in
coverage estimates, reflecting the limited statistical stability of ensemble diagnostics in 
a high-dimensional parameter space rather than inadequate sampling of individual regions of that space. Each test spectrum was treated as a mock observation and analysed using the
emulator-based MCMC framework, allowing us to compare the recovered posteriors with the known
“true’’ parameter values.

\subsection{Metrics}

For each parameter $\theta$\ we compute: (i) the mean bias between the recovered posterior mean and the true value, expressed both in physical units (bias) and normalised by the posterior standard deviation ($\mathrm{bias}_z$), (ii) the standard deviation of the probit-transformed probability integral transform values ($s_z$), (iii) the root-mean-square error (RMSE) between recovered means and true values in physical units, and (iv) the empirical coverage fractions of the 68\% and 95\% highest posterior density (HPD) intervals. These quantities jointly assess accuracy (bias, RMSE) and the reliability of the quoted uncertainties (coverage). We note that $\mathrm{bias}_z$ measures the mean offset between recovered and true values in units of the quoted posterior uncertainty, and therefore provides a dimensionless diagnostic of systematic trends relative to the inferred uncertainty scale, rather than a direct measure of physically significant offsets.

The probability integral transform (PIT) for each parameter and validation spectrum is defined as $\mathrm{PIT}=F_{\theta}(\theta_{\rm true})$, where $F_{\theta}$ is the marginal posterior cumulative distribution function predicted by the emulator and $\theta_{\rm true}$ is the known input value. Under correct specification, PIT values are independent and uniformly distributed on [0,1]. For visual and quantitative diagnostics we use their normal scores,
\[
z = \Phi^{-1}\!\big(\mathrm{PIT}\big),
\]
where $\Phi^{-1}$ is the inverse of the standard normal cumulative distribution function (CDF), commonly referred to as the probit function.
For a well-calibrated predictive model, these normal scores $z$ should follow a standard normal distribution $\mathcal{N}(0,1)$ \citep{gneiting2007, modrak2022}. In this formulation, an unbiased and properly scaled emulator yields $\mathrm{bias}_z \approx 0$, $s_z \approx 1$, and empirical coverages close to the nominal 68\% and 95\% levels.

\begin{figure*}[h]
  \centering
  \includegraphics[width=\textwidth]{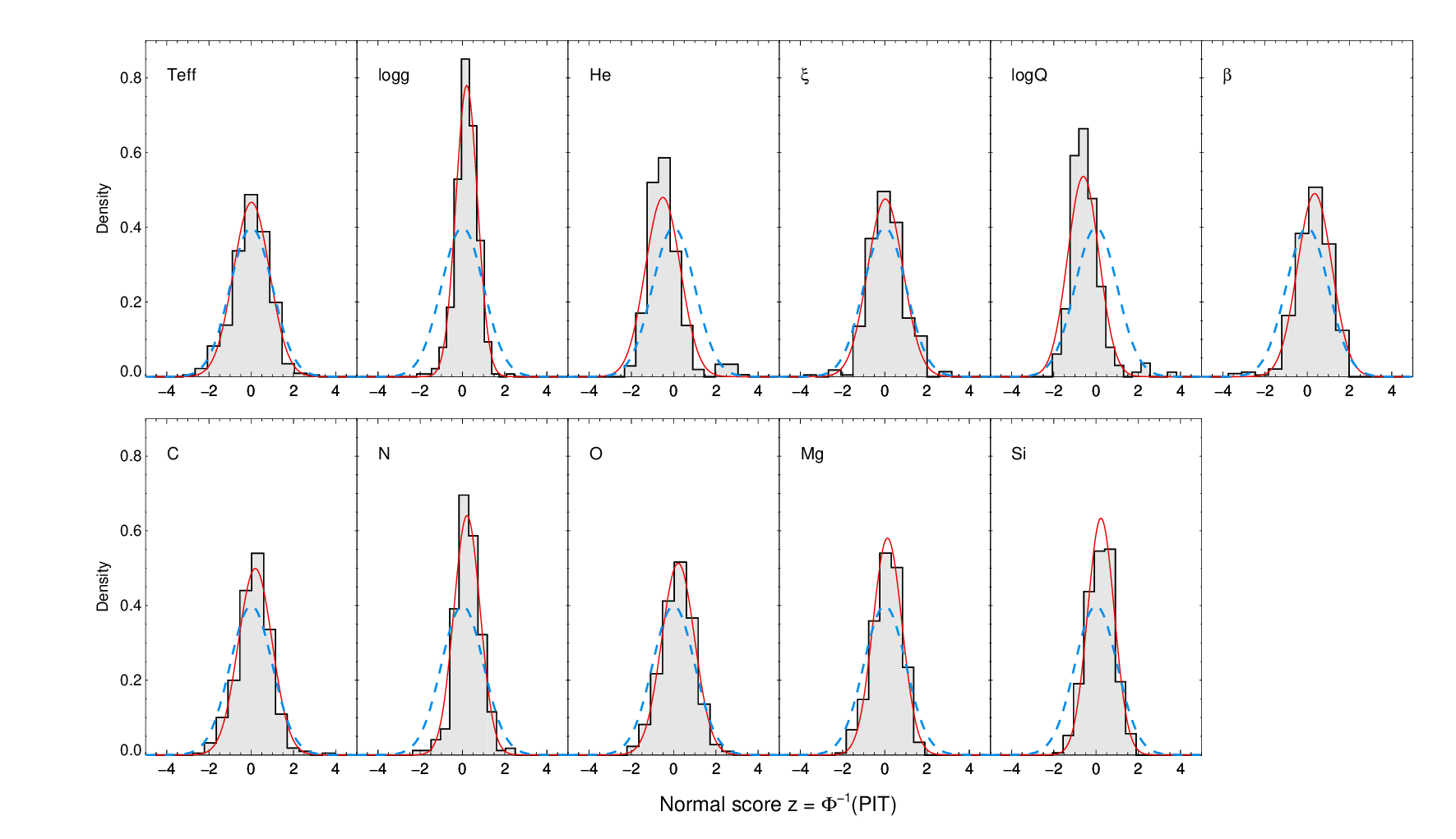}  
\caption{Calibration diagnostics using normal scores of the PIT. The panels show the empirical distributions of 
\(z\) for the difference parameters defining the emulator. The blue dashed curve marks the standard normal \(\mathcal{N}(0,1)\), with the red
solid line providing a kernel density estimate of the results.
}
  \label{fig:validation_zscores}
\end{figure*}

Figure~\ref{fig:validation_zscores} illustrates the distributions of the normal scores $z$ 
compared with a standard normal (blue dashed line). The close agreement indicates that the emulator is unbiased and well calibrated, with only modest deviations for a few parameters. Together with the summary statistics in Table~\ref{tab:emulator_validation}, this demonstrates that the emulator reproduces the correct posterior distributions within the expected uncertainties.

\subsection{Implications}
Overall, the validation results show that the emulator performs reliably across all 11 parameters. For photospheric quantities such as \teff\ and \logg, systematic biases are negligible, remaining below $\sim$10~K and $\sim$0.01~dex, respectively, while absolute errors are small, with RMSE values of order $\sim$170~K in \teff\ and $\sim$0.02~dex in \logg. 
The chemical abundances of metals, He, and the microturbulent velocity likewise show null biases and low RMSE values (typically below 0.05 dex or their equivalent), with any larger normalised biases reflecting the small inferred posterior uncertainties rather than physically significant offsets. Such small offsets are not expected to produce any noticeable effect on the predicted spectra and thus confirm the accurate recovery of these parameters. The emulator thus provides a
faithful and efficient surrogate of the underlying models. We note that the recovery of wind parameters 
is intrinsically more challenging in the optical,
where H$\alpha$ becomes less sensitive at low mass-loss rates. As a result, both $\log Q$ and $\beta$
tend to show broader, more degenerate posteriors in this regime. This reflects the limitations of
the diagnostic rather than shortcomings of the emulator itself.

The outcome of this validation exercise is clear: across the parameter space covered by the
training grid, the emulator reproduces the results of the forward simulations within the quoted
uncertainties. The posterior distributions obtained with the e\-mu\-la\-tor show consistent coverage,
with the 68\% intervals tending to be slightly conservative. This implies that the emulator is, if
anything, underconfident rather than overconfident in its uncertainty estimates. This conclusion 
is supported by the scaling factors listed in the last column of Table~\ref{tab:emulator_validation}, all 
of which are smaller than unity, indicating that the quoted errors would need to be reduced to 
reach perfect calibration. Such mild underconfidence is a desirable feature, as it avoids the risk 
of underestimated uncertainties and ensures robustness when applied to real data.

\begin{table*}
\centering
\caption{Validation statistics for the emulator based on 400 simulations.} 
\small
\begin{tabular}{
l
S[table-format=+1.3]
S[table-format=1.3]
S[table-format=1.2]
S[table-format=1.2]
S[table-format=3.3]
S[table-format=+2.2]
S[table-format=1.2]
}
\hline\hline
Parameter & {$\mathrm{bias}_z$} & {$s_z$} & {cov68} & {cov95} & {RMSE} & {bias} & {err\_scl} \\
\hline
\teff\ [K]    &  0.012 & 0.854 & 0.75 & 0.98 & 164.510 &   8.35 & 0.77 \\
\logg\ [dex]              &  0.203 & 0.512 & 0.92 & 1.00 &   0.019 &  -0.01 & 0.47 \\
He/H                           & -0.510 & 0.831 & 0.73 & 0.98 &   0.007 &   0.00 & 0.62 \\
$\xi$ [km\,s$^{-1}$]     &  0.024 & 0.838 & 0.78 & 0.99 &   0.212 &  -0.03 & 0.75 \\
$\log Q$ [dex]             & -0.600 & 0.744 & 0.81 & 0.98 &   0.128 &   0.09 & 0.28 \\
$\beta$                       &  0.353 & 0.814 & 0.62 & 0.99 &   0.157 &  -0.07 & 0.82 \\
C [dex]                        &  0.191 & 0.799 & 0.79 & 0.99 &   0.030 &  -0.00 & 0.72 \\
N [dex]                        &  0.217 & 0.623 & 0.86 & 1.00 &   0.039 &  -0.01 & 0.57 \\
O [dex]                        &  0.217 & 0.777 & 0.77 & 0.99 &   0.025 &  -0.01 & 0.71 \\
Mg [dex]                      &  0.124 & 0.687 & 0.84 & 1.00 &   0.024 &  -0.00 & 0.63 \\
Si [dex]                       &  0.234 & 0.629 & 0.75 & 0.99 &   0.038 &  -0.01 & 0.73 \\
\hline
\end{tabular}
\tablefoot{Columns give the mean bias in $z$ ($\mathrm{bias}_z$), 
standard deviation of the probability integral transform (PIT) values ($s_z$), 
the coverage fractions at 68\% and 95\%, the RMSE of the recovered parameters, 
the bias in parameter space, and the error-scaling factor. 
Units of RMSE and bias are given in the native units of each parameter.}
\label{tab:emulator_validation}
\end{table*}

\section{Analysis of benchmark stars}
\label{sec:benchmark}
The results of the previous section demonstrate that our framework can reliably recover
input parameters from synthetic data. We now turn to the crucial step of applying the method
to real observations, using well-studied stars as benchmarks to evaluate its performance in
practical astrophysical settings. We analyse a sample of Galactic OB-type stars with
high-quality spectroscopic data available from public databases, many of which have published
atmospheric parameters and abundances based on classical quantitative analyses. 
The study of such well-characterised stars provides an effective reference set for assessing our
methodology in comparison with literature results. Moreover, the relatively simple atmospheres
of OB dwarfs and giants offer an ideal environment to test the reliability of the underlying atomic
data, without the additional complications introduced by strong stellar winds or chemical
peculiarities.  

In the following subsections, we briefly describe the observational data and selected spectral
diagnostics, outline the main assumptions of our analysis, present the results of the inference
process, and finally compare them with values reported in the literature.

\subsection{Observational data}

The benchmark sample was designed to span a representative range of late O- and early B-type 
dwarfs and giants, with low projected rotational velocities to minimise line blending and to enable 
stringent tests of the adopted model atoms. It comprises 27 Galactic OB stars covering spectral 
types O9 to B3.  
Most spectra were retrieved 
from the IACOB database \citep{simondiaz2011a,simondiaz2015}, complemented by a few additional 
cases from the Melchiors database \citep{royer2024}. The observations have high resolving power 
($R = 25000$--$85000$) and signal-to-noise ratios above $S/N \sim 200$ per pixel, 
ensuring that all diagnostic lines are well measured. The wavelength coverage extends from 
3800--7000~\AA, and up to 9200~\AA\ for the data collected with the HERMES spectrograph. 
Table~\ref{tab:targets} summarises the basic stellar and observational information.

For consistency and clarity, we focus primarily on stars previously analysed using classical methods in 
combination with \fastwind~models,  
aiming to isolate differences that arise from the inference methodology rather than from the underlying 
model-atmosphere code. Minor discrepancies may nonetheless reflect updates in the atomic data, as 
some of the present model atoms differ from those adopted in earlier studies.
To provide a broader basis for future cross-comparisons, we also include a supplementary set of bright, 
apparently normal OB-type dwarfs and giants that have not been subject to detailed 
quantitative analysis with \fastwind. These stars were selected for their high-quality spectra, 
low projected rotational velocities, and complementary positions in terms of 
ionisation balance and line-strength regimes. Throughout, we distinguish between 
the literature comparison sample and the supplementary set, which serves to test 
the robustness of the methodology and mitigate potential selection biases.

\subsection{Diagnostic spectral features}
\label{sec:lines}
The optical spectra of OB dwarfs contain a variety of diagnostic lines that allow us to 
constrain the fundamental parameters and abundances. The diagnostic set includes 
Balmer lines from H$\epsilon$ to H$\alpha$, selected He~{\sc i} and He~{\sc ii} 
transitions (e.g., He~{\sc i} 4471, 4922; He~{\sc ii} 4541, 5411), and representative metal 
lines from C, N, O, Mg, and Si. 
A complete list of all transitions considered is provided in Appendix~\ref{app:lines}. Together, 
these lines constitute a well-balanced diagnostic set, covering multiple ionisation stages and 
a range of line strengths. Combined with the inference algorithm described above and the 
assumptions outlined in the following section, they form the foundation of our quantitative 
spectroscopy framework.

\subsection{Fundamental assumptions}
\label{sec:assumptions}

Our analysis follows the principle of minimising prior assumptions, ensuring that the 
observational data provide the primary constraints on the inferred parameters. While fixing certain 
quantities to expected values can be a pragmatic choice in many studies, such 
assumptions may also 
reduce sensitivity to astrophysically relevant inconsistencies. 
For example, apparent abundance anomalies might arise from spectral contamination from
undetected companions, or intrinsic peculiarities of the star. By treating 
abundances and other parameters as free quantities, we ensure that the 
inference remains sensitive to such signals, allowing the data to reveal 
departures from expectation rather than enforcing them a priori.

Our analysis relies on several key assumptions, some of them departing 
from classical spectros\-copic studies:

\begin{enumerate}
\item Simultaneous multi-line fitting: All diagnostic lines of H, He, 
and metals are fitted simultaneously within a single Bayesian framework. This 
naturally accounts for correlations between parameters without requiring 
separate two-step determinations.

\item Microturbulence: A value of $\xi$ is required in the 
atmosphere calculations, where it can in principle influence the level 
populations through line opacities and radiative rates, while in the formal 
solution it acts as an additional Doppler broadening term. In constructing 
the training grid we adopted a fixed $\xi = 10~\mathrm{km\,s^{-1}}$ for the model
atmosphere, whereas in 
the inference process $\xi$ is treated as a free parameter shaping the line 
profiles. A detailed discussion of how microturbulence is constrained in our 
approach is given in Sect.~\ref{sec:micro_and_mcmc}.

\item Wind properties: For low luminosity OB stars, winds are weak. We model 
the wind-strength parameter $Q$ assuming smooth, unclumped outflows, 
acknowledging that small clumping effects may be present but are unlikely to 
significantly affect the optical lines used in this study.

\item Rotational and macroturbulent broadening: For each star we adopt 
initial estimates of $v \sin i$ and radial--tangential macroturbulence 
from the literature, incorporated as Gaussian priors 
with means and standard deviations reflecting the published values.

\item Abundances: All elemental abundances are treated as free 
parameters and constrained simultaneously, rather than adopting {\it expected} 
solar values or fixed ratios.

\end{enumerate}

\subsection{Results} 
\label{sec:results}

Posterior distributions for each star were derived using the Bayesian inference framework outlined in Sect.~\ref{sec:mcmc}. 
The resulting estimates provide robust stellar and wind parameters with fully quantified uncertainties and covariances. The 
fundamental parameters are summarised in Table~\ref{tab:fundamental_params}, and the corresponding chemical abundances 
in Table~\ref{tab:abundances}. Medians and 68\% credible intervals are reported throughout. Together, these results provide a 
complete quantitative description of the benchmark Galactic OB stars analysed in this work.

\begin{figure*}[ht]
\sidecaption
\includegraphics[width=12cm]{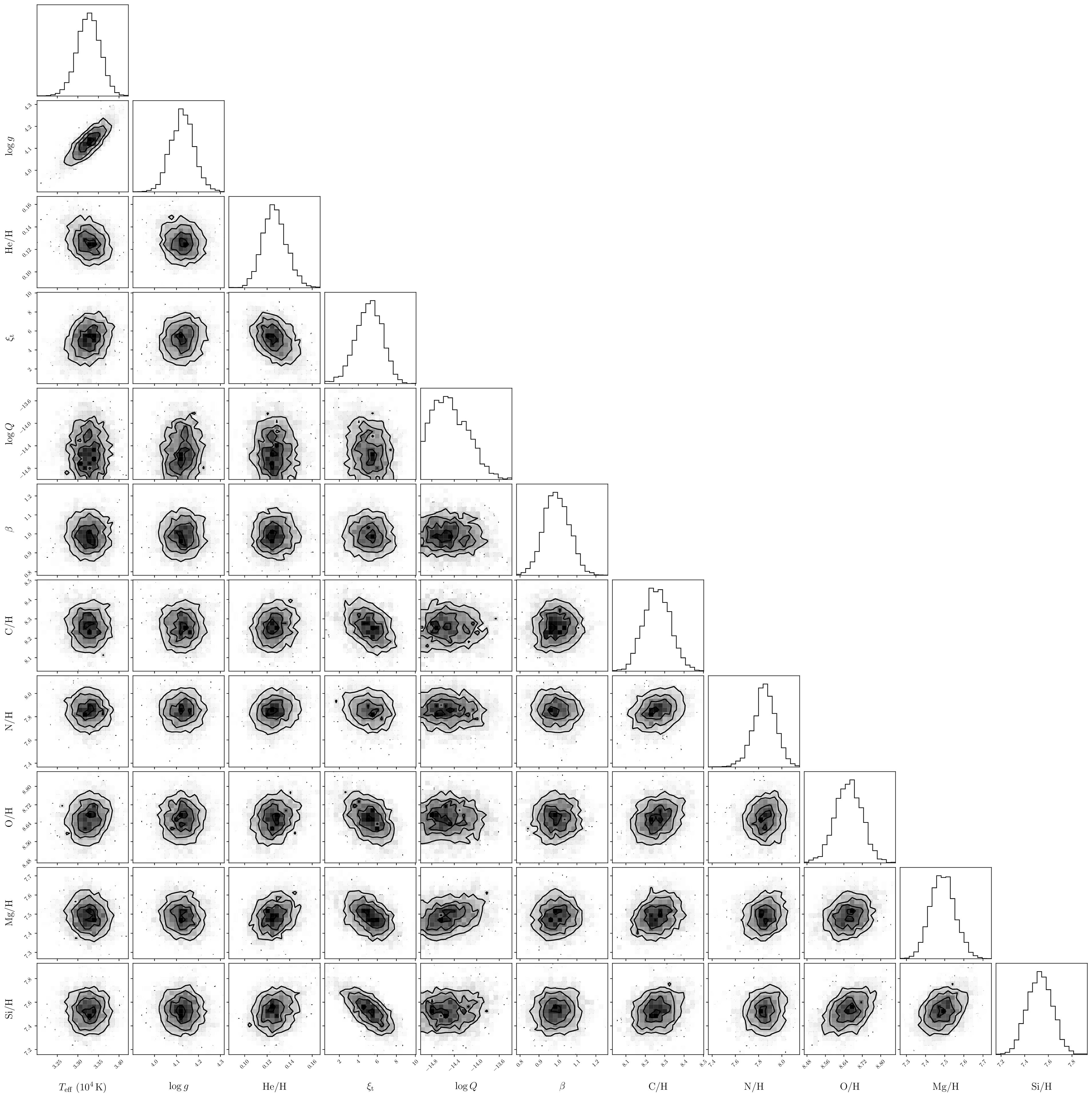}  
\caption{Posterior distributions for HD\,36512 (O9.7\,V). 
The results illustrate the typical parameter correlations obtained in our analysis: 
a clear degeneracy between \teff\ and \logg, and between $\xi$ and the He and metal abundances. 
Among the latter, the correlation with Si is strongest, reflecting the sensitivity of the Si~{\sc iii} triplet to 
microturbulence.}
  \label{fig:benchmark_corner}
\end{figure*}

Figure~\ref{fig:benchmark_corner} shows a representative corner plot for HD\,36512, illustrating both the marginal posterior 
distributions and the covariances among parameters. Correlations such as that between microturbulence and abundances, or 
between \teff\ and \logg, emerge naturally from the joint posterior. This exemplifies a key advantage of the 
Bayesian framework: parameter degeneracies are explicitly quantified, ensuring that the reported uncertainties reflect the true structure of the solution space.

For the wind parameters, the posteriors of $\log Q$ and $\beta$ show the limited sensitivity of optical diagnostics to weak winds.
This behaviour is fully consistent with expectations for late-O and early-B dwarfs and giants with weak winds. In this regime H$\alpha$ remains predominantly photospheric, with only subtle wind filling in the line core, so the optical diagnostics are mainly sensitive to the overall wind-density scaling encoded in $\log Q$, and carry very little direct information on the detailed shape of the velocity law. As a consequence, the posteriors for $\beta$ are largely prior-dominated  
and should be interpreted as reflecting the limited sensitivity of the data. In practice, our results provide robust upper limits and loose constraints on the wind strength, but essentially no meaningful constraint on $\beta$ for most stars in the sample.

Figures~\ref{fig:o_benchmark_fits} and~\ref{fig:b_benchmark_fits} illustrate the overall quality of the results for the O- and B-star subsamples, respectively. The comparisons across broad spectral windows show that the tailored \fastwind\ simulations computed using the posterior parameters accurately predict the H, He, and metal lines simultaneously for multiple stars, underscoring the robustness and internal consistency of the Bayesian analysis. Figures~\ref{fig:full_hd34078} and~\ref{fig:full_hd36591} present extended spectral ranges for two representative objects, HD~34078 (O9.5\,V) and HD~36591 (B1\,V). The close agreement across hydrogen, helium, and metal lines demonstrates that the derived parameters provide a consistent description of individual spectra—not only for the diagnostic lines explicitly used in the likelihood, but across the full optical range.

\subsubsection{Abundances}
\begin{figure}[ht]
\resizebox{\hsize}{!}{\includegraphics{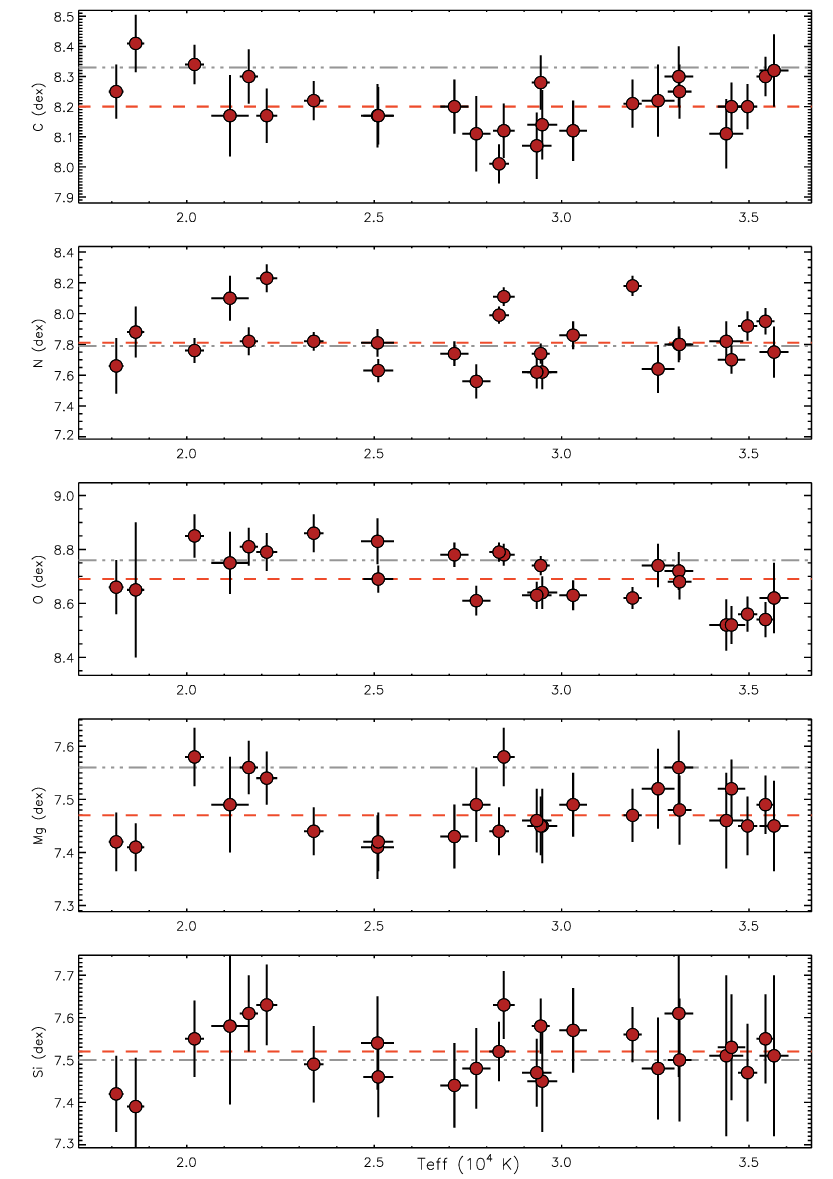}}   
\caption{Trends of the derived metal abundances with effective temperature. Each panel shows the inferred abundances of C, N, O, Mg, and Si as a function of \teff\ for the analysed sample. The error bars correspond to the 68\% credible intervals from the posterior distributions. A mild decrease in O abundance toward higher \teff\ is visible, consistent with the limitations of the current O\,\textsc{ii} model atom at high ionisation stages (see text); no significant trends are found for the other elements. The dash-dotted line marks the CAS value, and the dashed line indicates the median of the derived abundances in each panel.}
 \label{fig:abundances}
\end{figure}

A mild decrease in the inferred oxygen abundance with increasing \teff\ is apparent across the sample (see Fig.~\ref{fig:abundances}). 
The hottest objects tend to show slightly lower abundances than the cooler B-type stars. We interpret this trend as most likely originating from 
limitations in our current O\,\textsc{ii} model atom at higher ionisation stages rather than as evidence for a genuine depletion of oxygen at higher 
\teff. An updated O\,\textsc{ii} model atom incorporating improved atomic data and extended level structure should help to clarify this issue in future analyses. By contrast, neither Mg nor Si shows any significant dependence on \teff\ or \logg. Both elements remain approximately constant within the quoted uncertainties over the full range of stellar parameters sampled. 

To assess the absolute abundance scale, we compared the distribution of our inferred elemental abundances with the Cosmic 
Abundance Standard (CAS) derived for nearby early-type stars by \citet{przybilla2008} and \citet{nieva2012}. For Mg and Si, the median 
values in our sample are 
consistent with the CAS within the 1$\sigma$ scatter, with a mild systematic difference for Mg (median difference of 0.06 dex). Likewise, the median C abundance shows a shift of 0.12\,dex, with our derived values being lower. Nitrogen abundances show a consistent 
agreement to the CAS baseline, although slightly higher in our case (0.08 dex). 
Oxygen is on average consistent with the CAS value by $\sim$0.06~dex, albeit with the aforementioned mild \teff\ trend discussed above.  
Overall, the absolute scale of the inferred abundances is compatible with the CAS, and departures from it follow physically interpretable patterns—N enrichment and a possible O offset in the hottest stars—rather than random star-to-star variations. For reference, the Cosmic Abundance Standard itself exhibits a star-to-star dispersion of about 0.05–0.10~dex for C, N, and O. The small differences found here therefore lie well within the intrinsic scatter of the CAS sample and are consistent with normal Galactic abundance variations at solar metallicity.

No systematic trend is found between the derived abundances and the inferred microturbulent velocities, confirming that the adopted treatment 
of $\xi$ does not introduce spurious correlations in the abundance determinations (see below).

In summary, the abundance patterns inferred by \maui\ satisfy two key physical expectations. First, Mg and Si remain approximately constant and do not correlate with \teff\ or \logg, supporting their internal consistency and the high quality of the model atoms. Second, the absolute abundance scale is broadly consistent with CAS values, with a plausible modelling-driven offset in oxygen at the highest \teff. 
Taken together, these results show that the abundances delivered by the Bayesian multi-line analysis are not only statistically well constrained, but also astrophysically sound.

\subsection{How is the microturbulence constrained?}
\label{sec:micro_and_mcmc}

From a spectroscopist’s point of view, microturbulence is a phenomenological 
parameter introduced to remove systematic trends of derived abundances with 
line strength, ensuring consistency between weak and strong lines of the same 
species \citep[e.g.][]{struve1929, struve1934, gray2005}. Despite its empirical nature, 
it plays a crucial role in quantitative spectroscopy. Historically, different 
strategies have been adopted. Some studies derived species-dependent values of 
$\xi$  \citep[e.g.][]{trundle2004, simondiaz2010, carneiro2019}, while in the specific case
of B-type stars, different authors relied either on O~{\sc ii} lines from different multiplets 
\citep[e.g.,][]{kilianmontenbruck1994, korn2000}, or exclusively on the Si~{\sc iii} triplet  
lines 4553-67-74 \citep[e.g.,][]{kilian1992, mcerlean1998, urbaneja05b, urbaneja11}, exploiting 
its wide range of equivalent widths within a single multiplet to minimise the effect of 
modelling uncertainties. 
These approaches have proven effective in practice, though they may yield 
species-dependent results and make it less straightforward to propagate 
uncertainties consistently into other parameters. 
Attempts have also been made to link the spectroscopic $\xi$ to underlying 
physical processes, most notably sub-surface convection associated with the 
iron opacity peak \citep{cantiello09}. While such works provide valuable 
insight into a possible origin, the relation between the empirically derived 
values and the actual velocity fields in stellar atmospheres remains uncertain.

\begin{figure*}[!ht]
  \centering
  \includegraphics[width=0.8\textwidth]{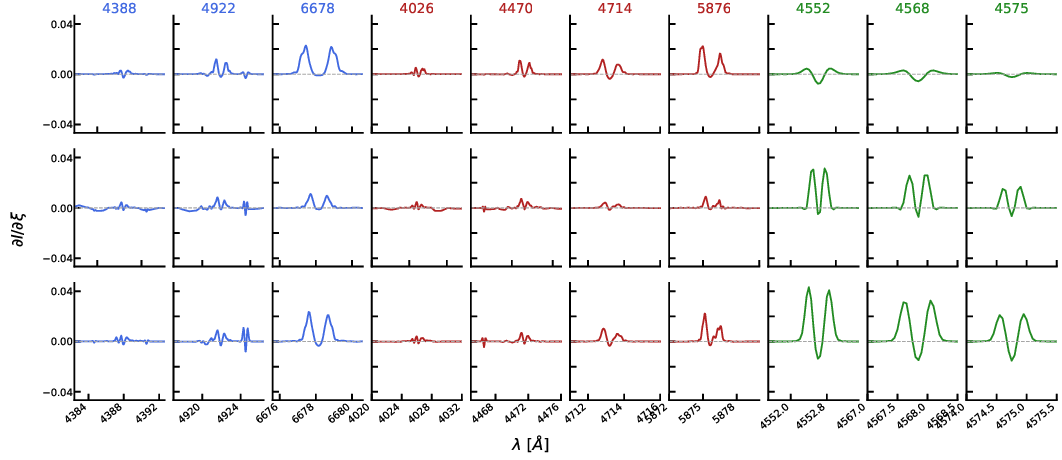}  
\caption{Sensitivity of selected diagnostic lines to microturbulence, shown as 
the differential response of the emergent spectrum to changes in $\xi$. Each 
row corresponds to a dwarf star of spectral type O9, B0, and B1, respectively. 
The first seven panels show He\,\textsc{i} singlet (4387, 4922, 6678\,\AA) and 
triplet (4026, 4471, 4713, 5876\,\AA) lines, followed by the three components 
of the strong Si\,\textsc{iii} triplet (4553, 4568, 4575\,\AA). }
  \label{fig:he_and_micro}
\end{figure*}

We adopt a complementary strategy in which a single, global $\xi$ is treated as a free parameter 
in the Bayesian inference, constrained simultaneously by lines of different species and strengths. 
Figure~\ref{fig:he_and_micro} illustrates the rationale behind this approach. 
It shows the sensitivity of selected He\,\textsc{i} lines and of the Si\,\textsc{iii} triplet to changes in $\xi$, 
for three representative sets of parameters corresponding to generic O9 (top), B0 (middle), and B1.5 (bottom) dwarfs. 
Within the Si\,\textsc{iii} triplet, the stronger $\lambda4553$ line reacts
more strongly than the weaker $\lambda4574$ component, with $\lambda4567$ showing an intermediate behaviour.
This differential sensitivity within a single multiplet is precisely what led to the widespread use of these lines 
to constrain $\xi$ robustly in B stars. By contrast, in the hotter O9 case the same lines are weak, showing an almost flat
response to  $\xi$, largely removing their diagnostic power. In this domain, constraining microturbulence must 
therefore rely on alternative diagnostics --other ions or line sets-- that remain sensitive to $\xi$. 
These differences emphasise that the effect of microturbulence depends on the detailed line-formation conditions 
and on the location of the star in parameter space.

Such behaviour underscores the need for a global treatment of $\xi$. 
Because the diagnostic sensitivity varies with spectral types, no single line set remains informative 
throughout the parameter space. 
By combining all available diagnostics within a unified Bayesian framework, 
our method naturally employs the lines that are most sensitive under the local physical conditions, 
allowing microturbulence to be constrained consistently from the information that is actually available in each regime. 

The posterior constraints (Fig.~\ref{fig:benchmark_corner}) confirm that $\xi$ is well constrained by the combined diagnostics. 
Both weak and strong lines of individual elements are reproduced simultaneously in our tailored models 
(Figs.~\ref{fig:o_benchmark_fits}--\ref{fig:full_hd36591}), providing empirical support for the inferred microturbulent velocities. 
This agreement across different parts of the curve of growth indicates that the adopted $\xi$ values re\-con\-ci\-le abundance 
determinations from lines of different strengths in a statistically consistent way. 
By integrating all available diagnostics within a single Bayesian framework, our approach mitigates the biases 
inherent to line- or species-specific determinations and anchors the inferred microturbulence to the full range of evidence.
Finally, we want to note that $\xi$ is adjusted only in the formal solution, not in the atmospheric structure itself, 
which may introduce small internal inconsistencies; these are expected to have only minor impact on the present analysis.

Concerning our derived values, a mild systematic behaviour of $\xi$ is apparent within the sample:  
among the dwarfs, there is a tentative indication that $\xi$ increases slightly with effective temperature, although the effect remains within the formal uncertainties. This tendency is qualitatively consistent with previous findings for OB stars and likely reflect the combined influence of atmospheric extension and thermal structure, but confirmation will require a larger and more homogeneous sample.

\subsection{Comparison with previous studies}
\label{sec:literature}

We compared our results with published \fastwind\ based analyses in order to minimise code--dependent 
systematics and to isolate differences arising from (i) methodology, (ii) observational material, and (iii) atomic data.

\paragraph{O-type stars}
For the late-O stars we used as references the sequence of \fastwind\ analyses by \citet{holgado18}, \citet{carneiro2019}, \citet{martinezsebastian2025}, and the most recent work by \citet{holgado2025}. These studies share a common methodology and provide an internally consistent reference set for assessing our Bayesian inference approach in this hot regime. While all of them deliver fundamental parameters, detailed metal abundances are only provided by \citet[][, C, N, O]{carneiro2019} and \citet[][, N]{martinezsebastian2025}.

Across the three hottest objects (HD\,214680, HD\,46202, and HD\,34078), our derived effective temperatures and surface gravities agree closely with those of \citet{carneiro2019}, with typical differences well within the combined 1$\sigma$ uncertainties. Median offsets are of order 70~K in \teff\ and 0.04~dex in \logg, fully consistent with expected variations from diagnostic weighting. HD\,36512 shows a larger difference in \logg\ and He abundance. In contrast, the recent analysis by \citet{martinezsebastian2025}, which follows the same methodology as \citet{carneiro2019}, yields parameters consistent with ours—including the N and elevated He abundances. This suggests that the discrepancy with respect to \citet{carneiro2019} most likely originates from differences in the observational material or its reduction rather than from the inference technique itself.

Microturbulent velocities agree to within $\pm0.5$~km\,s$^{-1}$ for all stars in common. Carbon and nitrogen abundances are consistent within uncertainties (mean differences of $-0.05\pm0.10$ and $+0.03\pm0.12$~dex, respectively), while oxygen is systematically higher in our analysis by $\simeq0.2$~dex, consistent with the caution raised by \citet{carneiro2019} regarding the limitations of their adopted oxygen model atom.

The formal uncertainties in \teff\ and \logg\ are of the same order as those reported in previous works, reflecting that these parameters are already tightly constrained by well-established diagnostics such as Balmer-line wings and ionisation equilibria. In contrast, the chemical abundances show a clear gain in precision: our formal errors in C and O are smaller by factors of two to five, depending on the star, whereas those in N are comparable or slightly larger. This improvement stems from the global, multi-line nature of the Bayesian inference, which exploits all available spectral information and, by marginalising over all parameters simultaneously, properly accounts for covariances between \teff, \logg, $\xi$, 
and the elemental abundances. As a result, the derived abundance uncertainties are more realistic and statistically homogeneous across the sample.

To extend the comparison in this regime, we also considered four
additional O-type stars (HD\,44597, HD\,161789, HD\,166546, and
HD\,216898) recently analysed by \citet{holgado2025}, who provide \teff, \logg, and He abundances. 
The effective temperatures agree within 0.3$\sigma$ for all stars, with a median offset of 0.2$\sigma$. Our formal uncertainties are of the same order as those quoted by \citet{holgado2025}. Surface 
gravities show similarly close agreement, with a median difference of 0.3$\sigma$ and a maximum of 1.1$\sigma$, again 
accompanied by uncertainties smaller by about a factor of two. Helium
abundances are consistent in all four cases. The median ratio of He uncertainties is $\sim$0.6, confirming 
that our inference yields slightly tighter but statistically compatible constraints.

\paragraph{B-type stars}

For the B-type stars we compared our results with stars in the Orion OB1 sample analysed by \citet{simondiaz2010}, where \teff\ is determined from 
the Si\,\textsc{ii/iii/iv} ionisation balance. The sample includes HD~36512, HD~36960, HD~36591, HD~36959, HD~35299, HD~35039, 
and HD~36430. These objects provide a homogeneous and well-studied reference set for assessing the performance of our Bayesian inference at cooler effective temperatures.

Across the sample, the fundamental parameters show close agreement
with the literature. For both $T_{\mathrm{eff}}$ and $\log g$, the
typical differences correspond to about 0.4 of the combined 1$\sigma$
uncertainty, with the largest deviations remaining below 1.2$\sigma$
and 1.8$\sigma$, respectively. Our formal 1$\sigma$ uncertainties are
typically 300~K in \teff\ and 0.05~dex in $\log g$, compared to $\sim$500~K and 0.10~dex in the reference study. This corresponds to uncertainty ratios of roughly 0.6 and 0.5, indicating that our analysis provides constraints of comparable or moderately higher precision for the stellar parameters.

The oxygen and silicon abundances are statistically consistent with those reported by \citet{simondiaz2010}. For oxygen, the differences between both studies are typically well within the combined 1$\sigma$ uncertainties, with an average offset of about half a sigma and the largest case reaching roughly 1.5$\sigma$. For silicon, the agreement is even tighter: the average deviation is only about 0.07~dex, and all stars lie within 1$\sigma$. The typical uncertainty ratios are about 0.7 for oxygen---our results being approximately 25\% more precise---and about 1.1 for silicon, indicating comparable precision overall. These comparisons confirm that the Bayesian multi-line inference yields abundances consistent with the literature while maintaining a uniform statistical treatment across elements.

The microturbulent velocities show somewhat larger star-to-star scatter. On average, the differences correspond to about one to one-and-a-half sigma, reaching up to three sigma in the most discrepant cases, while the typical ratio of formal uncertainties remains close to unity ($\approx$1.2). This suggests that both analyses assign similar formal errors and that the observed differences mainly reflect genuine stellar variations rather than mismatched error estimates. Such scatter is consistent with the known sensitivity of $\xi$ to the details of line formation, and small residual abundance offsets---particularly for oxygen and silicon in this case---are most likely attributable to differences in the underlying atomic data adopted in the respective analyses.

In addition to the main comparison sample, we included $\tau$~Sco (HD\,149438) as a reference object. This star is the prototypical B0\,V standard and has been the subject of numerous detailed spectroscopic analyses (e.g.\ \citealt{mokiem2005}; \citealt{simondiaz2006}). Its inclusion provides an additional anchor for the early-B regime and allows direct comparison with the extensive literature available for this benchmark. Moreover, $\tau$~Sco is a particularly intriguing case, as it is widely accepted—or at least strongly suspected—to be the product of a past stellar merger, which may explain its unusually high helium abundance and complex magnetic field structure. Including this object therefore tests the robustness of our inference scheme for stars with potentially non-standard surface compositions. The agreement of our inferred parameters with those obtained by \citet{mokiem2005} and \citet{simondiaz2006} is excellent. Unfortunately, none of these two studies determined metal abundances, hence a comparison is not possible.

For the B-type stars our \teff, \logg, O, and Si determinations are statistically consistent with the literature, typically within $<1\sigma$, and our formal uncertainties are of equal or smaller magnitude. The microturbulent velocities show larger object-to-object deviations at the $\sim$1–3$\sigma$ level, consistent with the expected methodological sensitivity of $\xi$. Overall, the results confirm that the Bayesian inference framework yields parameters and abundances fully compatible with previous analyses, while providing a homogeneous and statistically rigorous treatment across the B-star sample.\\

\begin{figure*}[ht]
\sidecaption
\includegraphics[width=12cm]{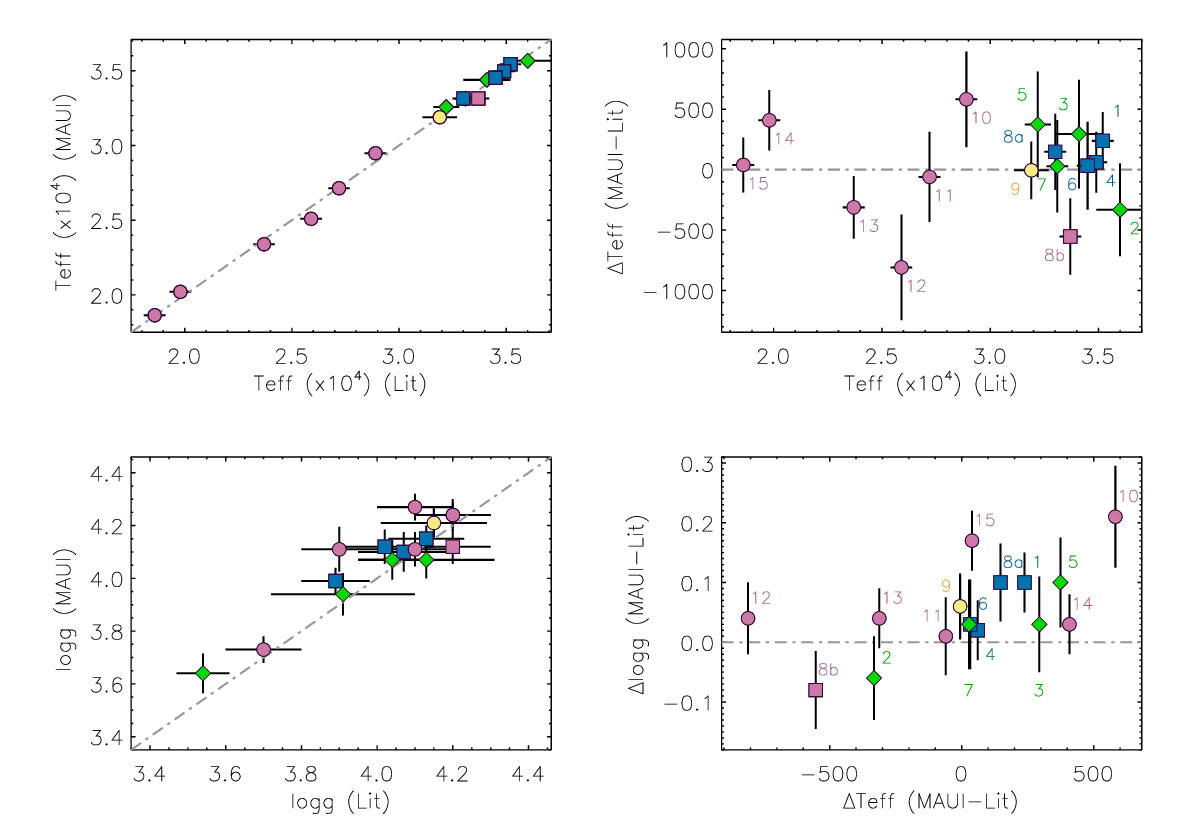}
\caption{Comparison with literature values. Late O stars: blue squares -- \citet{carneiro2019}, green diamonds -- \citet{holgado2025}; 
Early/mid B stars: yellow circle -- \citet[][, $\tau$ Sco]{mokiem2005}, purple circles -- \citet{simondiaz2010}. 
  \label{fig:litcomp} } 
\end{figure*}

Figure~\ref{fig:litcomp} presents the comparison for \teff\ and
\logg\ with the literature. O-type stars are shown as squares, with the corresponding
reference studies indicated by colour: blue—\citet{carneiro2019};
green—\citet{holgado2025}; pink—\citet{simondiaz2010}. The B-type
stars are displayed as circles: yellow—\citet{mokiem2005} ($\tau$~Sco); pink—\citet{simondiaz2010}.

For both the O- and B-type samples, our results are fully consistent with the stellar parameters and abundances obtained in previous \fastwind\ analyses within the quoted uncertainties, while providing smaller and more uniform formal errors. The improvements are most pronounced for 
\logg, \teff, and the CNO abundances. An exception is oxygen in the late-O stars, where our values are systematically higher than those of \citet{carneiro2019}. This difference is understood in light of the simplified oxygen model atom adopted in that study, as noted by the authors themselves. Overall, the remaining residuals are compatible with differences in atomic data or observational material.

\section{Discussion}
\label{sec:discussion}

The results presented above establish that \maui\ delivers stellar and wind parameters consistent with classical spectroscopic analyses while providing a full probabilistic characterisation of the uncertainties. By accounting for correlations among pa\-ra\-me\-ters and propagating uncertainties consistently, \maui\ goes beyond line-by-line approaches and offers a transparent quantification of the solution space. A guiding principle of our workflow is to minimise prior assumptions, allowing the observational data to provide the primary constraints: abundances, microturbulence, and other parameters are not fixed to expected values but are derived directly from the spectra. This ensures that potential anomalies are revealed by the analysis rather than masked by assumptions, while recognising that their origin may lie in the stars themselves, in the observations, or in limitations of the models and atomic data.

Although validated here on Galactic late-O and early-B stars, the framework is applicable to other stellar types and model grids. Applications in the literature already include optical spectroscopy of B- and A-type supergiants in the LMC \citep{urbaneja17}, near-IR spectroscopy of Galactic Cepheids \citep{inno2019}, and the large Galactic B-star study by \citet{deburgos2024}. These works confirm that MAUI is robust across spectral types, model-atmosphere codes, and wavelength regimes. A further advantage is computational efficiency: once trained, the emulator enables MCMC-based inference at a fraction of the cost of direct model-atmosphere calculations while retaining rigorous uncertainty propagation. Homogeneous a\-na\-ly\-ses of sizeable samples thus become tractable on standard computing resources.

Beyond the quantitative agreement with previous analyses and the
reduction in formal uncertainties, the Bayesian framework also carries an important conceptual implication for how stellar parameters are constrained from spectra.

Classical spectroscopic analyses traditionally associate each stellar parameter with a limited set of diagnostic lines. The effective temperature is derived from ionisation equilibria (for example, Si\,\textsc{ii/iii/iv}), the surface gravity from the wings of the Balmer and He\,\textsc{i/ii} lines, and the microturbulent velocity from abundance trends with line strength. This scheme is physically motivated and has long provided robust results, yet it implicitly treats the parameters as 
quasi-separable quantities, each constrained by an independent subset of observables.

In reality, the spectrum is a global manifestation of the same
underlying atmospheric structure. Every wavelength point originates
from a complex radiative-transfer calculation that depends on all
model parameters—temperature, density, velocity fields, and
composition. Consequently, changes in any parameter influence not only
its classical diagnostics but also many other spectral features. For instance, 
metal lines respond to variations in the density distribution and radiation field, thereby carrying indirect information on \logg; similarly, the thermal structure and line-blanketing effects couple the strengths of metal and helium lines to \teff. Mathematically, each point of the spectrum can therefore be regarded as sampling the multi-dimensional function $F_\lambda(\boldsymbol{\theta})$ that encodes the combined physics of the atmosphere, with varying sensitivity to each component of $\boldsymbol{\theta}$.

The Bayesian multi-line inference implemented in \maui\ na\-tu\-rally accounts for this coupling. By evaluating the likelihood across wide spectral ranges, the method automatically weighs each wavelength region according to its sensitivity to the parameters, given the model and data quality. Parameters are thus constrained jointly by all available information, without pre-selection of specific diagnostics. The familiar indicators continue to dominate where their sensitivities are strongest, but weaker correlations elsewhere contribute additional information and are reflected in the posterior covariances. This global treatment replaces the approximate separability of the classical approach with a statistically consistent description in which all parameters are inferred simultaneously from the full available information encoded in the spectrum.

The treatment of the microturbulent velocity as a single, global parameter exemplifies this principle: while $\xi$ has no unique diagnostic feature, its value influences the strength and shape of lines from multiple ions. In the Bayesian framework, $\xi$ is therefore constrained by the collective behaviour of all lines rather than by an isolated subset, just as every other parameter is informed by the full spectral information content.

The discussion above highlights that the strength of \maui\ lies not only in its quantitative 
performance but also in its physically consistent, global treatment of the information contained in stellar spectra.

\section{Closing remarks}
\label{sec:closing}

This work demonstrates that Gaussian-process emulators, combined with dimensionality reduction and MCMC sampling, 
provide a rigorous and computationally efficient framework for quantitative spectroscopy of massive stars. The approach yields transparent uncertainty estimates and propagates parameter covariances consistently, while reducing computational costs by orders of magnitude compared to direct atmosphere calculations. For the scale of analyses presented here, the method is fully tractable: a single MCMC run with $5\times10^{4}$ steps can be completed on standard workstations, enabling homogeneous studies of moderate-sized samples.

The methodology is broadly applicable to other stellar types and model grids. Extensions of the training grids, improved treatments of correlated GP components, and refined uncertainty calibration will further enhance the robustness of \maui\ for quantitative spectroscopy of hot stars. The current emulator is restricted to optical spectra, where wind parameters such as $\log Q$ and $\beta$ remain only weakly constrained; incorporating UV and IR diagnostics will be essential for a more complete description of stellar outflows.

The \maui\ framework is fully code-agnostic and can be coupled to any state-of-the-art stellar-atmosphere code. Its emulation capability is particularly advantageous for computationally intensive models such as CMFGEN \citep{hillier98}, where first tests have already been successfully conducted. The same strategy will enable systematic comparisons between different atmosphere codes across overlapping parameter regimes, and the inclusion of additional physical quantities—such as clumping parameters, including the treatment of optically thick clumping 
\citep[e.g.][]{hawcroft21, brands2022}—to investigate their correlations and impact on inferred stellar and wind properties.

For very large spectroscopic surveys, neural-network e\-mu\-la\-tors will
complement \maui\ by offering millisecond predictions at the cost of
less transparent uncertainty estimates. A hybrid strategy is therefore 
appealing: GP-based emulators such as \maui\ for rigorous, uncertainty-aware inference and validation, complemented by deep-learning surrogates for fast exploratory analyses. As stellar-atmosphere models evolve toward 3-D structures, time dependence, and detailed wind dynamics, their computational demands will continue to grow. Emulator-based in\-fe\-ren\-ce will thus become increasingly essential, with developments such as adaptive training, sparse or parallelised GPs, and systematic validation under realistic observational conditions ensuring that \maui\ remains accurate and robust across diverse spectral regimes.

By enabling robust and homogeneous stellar parameters, the framework directly supports applications ranging from stellar-evolution and feedback studies to the interpretation of large spectroscopic surveys. In this context, \maui\ provides a practical route toward exploiting the full information content of high-resolution spectra, bridging the gap between detailed atmosphere modelling and statistically rigorous inference.

In this way, emulator-based inference transforms quantitative spectroscopy from a labour-intensive art into a scalable, uncertainty-aware tool for the next generation of stellar astrophysics.

---

\begin{acknowledgements}
We are grateful to J. Puls for providing careful and insightful comments on the manuscript, which helped to improve its clarity and presentation. We also thank N. Przybilla for generously sharing his excellent model atoms, which were essential for this work. Finally, we thank the referee for a constructive and thoughtful report that helped to clarify several aspects of the paper.

The IACOB spectroscopic database is based on observations made with the Nordic Optical Telescope operated by the Nordic Optical Telescope Scientific Association, and the Mercator Telescope , operated by the Flemish Community, both at the Observatorio de El Roque de los Muchachos (La Palma, Spain) of the Instituto de Astrof\'{i}sica de Canarias.

\end{acknowledgements}

\bibliography{biblio}

\begin{appendix}

\onecolumn
\section{Benchmark stars}

\begin{table}[ht!]
\centering
\caption{Spectral classification and basic reference information for the reference set of stars.}
\label{tab:targets}
\begin{tabular}{lllccc}
\hline\hline
\# & Star & SpT & $R$ & Library & Literature \\
\hline
1  & HD214680 & O9V    & 46000 & 1  & 1  \\
2  & HD216898 & O9V    & 25000 & 1  & 2  \\
3  & HD44597  & O9.2V  & 85000 & 1  & 2  \\
4  & HD46202  & O9.2V  & 25000 & 1  & 1  \\
5  & HD166546 & O9.5IV & 46000 & 1  & 2  \\
6  & HD34078  & O9.5V  & 85000 & 1  & 1  \\
7  & HD161789 & O9.7IV & 25000 & 1  & 2  \\
8  & HD36512  & O9.7V  & 85000 & 1  & 1,3 \\
9  & HD149438 & B0V    & 85000 & 1  & 4  \\
10 & HD36960  & B0.5V  & 85000 & 1  & 3  \\
11 & HD36591  & B1V    & 46000 & 1  & 3  \\
12 & HD36959  & B1V    & 46000 & 1  & 3  \\
13 & HD35299  & B1.5V  & 46000 & 1  & 3  \\
14 & HD35039  & B2V    & 46000 & 1  & 3  \\
15 & HD36430  & B2V    & 46000 & 1  & 3  \\
16 & HD6675    & B0.2III & 46000 & 1 &    \\
17 & HD36822  & B0.2IV & 85000 & 1  &    \\
18 & HD25443  & B0.5III& 46000 & 1  &    \\
19 & HD218376 & B0.5III & 46000 & 1 & \\
20 & HD46328  & B0.7IV & 85000 & 1,2 &   \\
21 & HD201795 & B0.7V  & 85000 & 1  &    \\
22 & HD44743 & B1II-III & 46000 & 1 & \\
23 & HD66665  & B1V    & 25000 & 1  &    \\
24 & HD16582  & B2IV   & 85000 & 1  &    \\
25 & HD886    & B2IV   & 85000 & 1,2  &  \\
26 & HD35468  & B2III  & 85000 & 1,2  &  \\
27 & HD160762 & B3IV   & 85000 & 1,2  &  \\
\hline
\end{tabular}
\tablebib{
(1)~\citet[][]{carneiro2019}
(2)~\citet[][]{holgado2025}
(3)~\citet[][]{simondiaz2010}
(4)~\citet[][]{mokiem2005}
}
\tablefoot{Data source: (1) IACOB, (2) Melchiors.}
\end{table}

\FloatBarrier
\section{Diagnostic lines}

Table~\ref{app:lines} lists the spectral lines included in the likelihood evaluation. These features constitute the set of diagnostics used by 
\maui\ to constrain the stellar and wind parameters and chemical abundances in the Bayesian analysis. The adopted line selection 
covers the main ionisation equilibria of key elements, together with representative metal lines that are sensitive to 
microturbulence and abundance variations. 

\begin{figure*}[ht]
  \centering
  \includegraphics[width=0.7\textwidth]{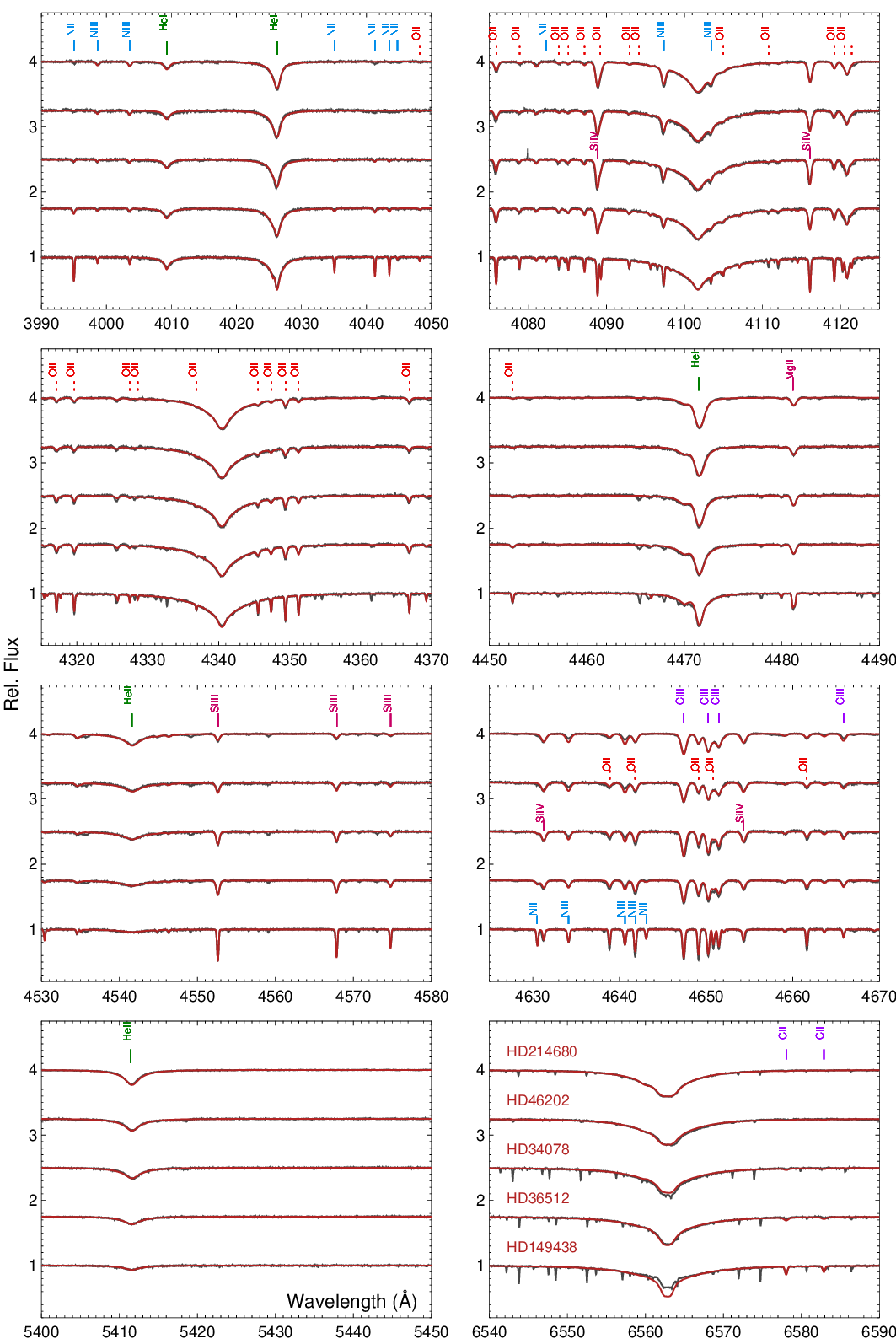}  
  \caption{
Representative spectral windows for selected O9-B0 stars. Each panel compares 
the observed spectra with the tailored \fastwind\ models 
calculated for the final parameters derived from the Bayesian analysis. 
Multiple stars are shown in each panel (vertically shifted for clarity). 
The simultaneous agreement across H, He, and metal lines demonstrates the 
robustness of the inferred parameters for the late O/early B stars. Due to the possible
effect of the magnetic field, the H$\alpha$ core was excluded from the evaluation of the 
model likelihood for HD\,149438. 
}
  \label{fig:o_benchmark_fits}
\end{figure*}

\begin{figure*}[ht]
  \centering
  \includegraphics[width=0.7\textwidth]{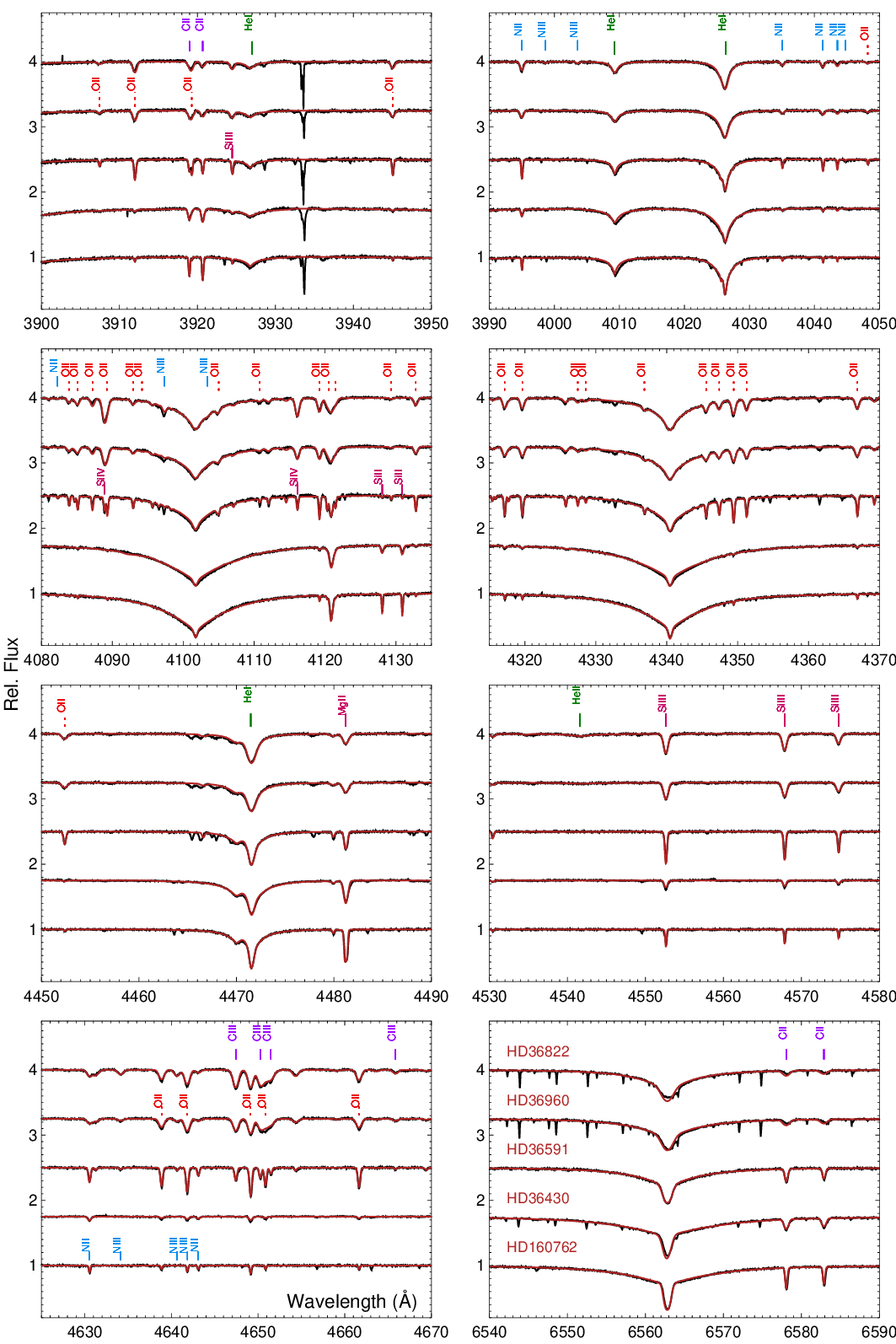}  
\caption{ As Fig.~\ref{fig:o_benchmark_fits}, but for selected B-type stars in the  
sample. The consistent 
fits across diagnostic lines of H, He, and metals highlight the reliability of 
the modelling and the internal consistency of the results for the B-type dwarfs/giants.
}
  \label{fig:b_benchmark_fits}
\end{figure*}

\begin{figure*}[ht]
 \centering
 \includegraphics[angle=90,width=0.7\textwidth]{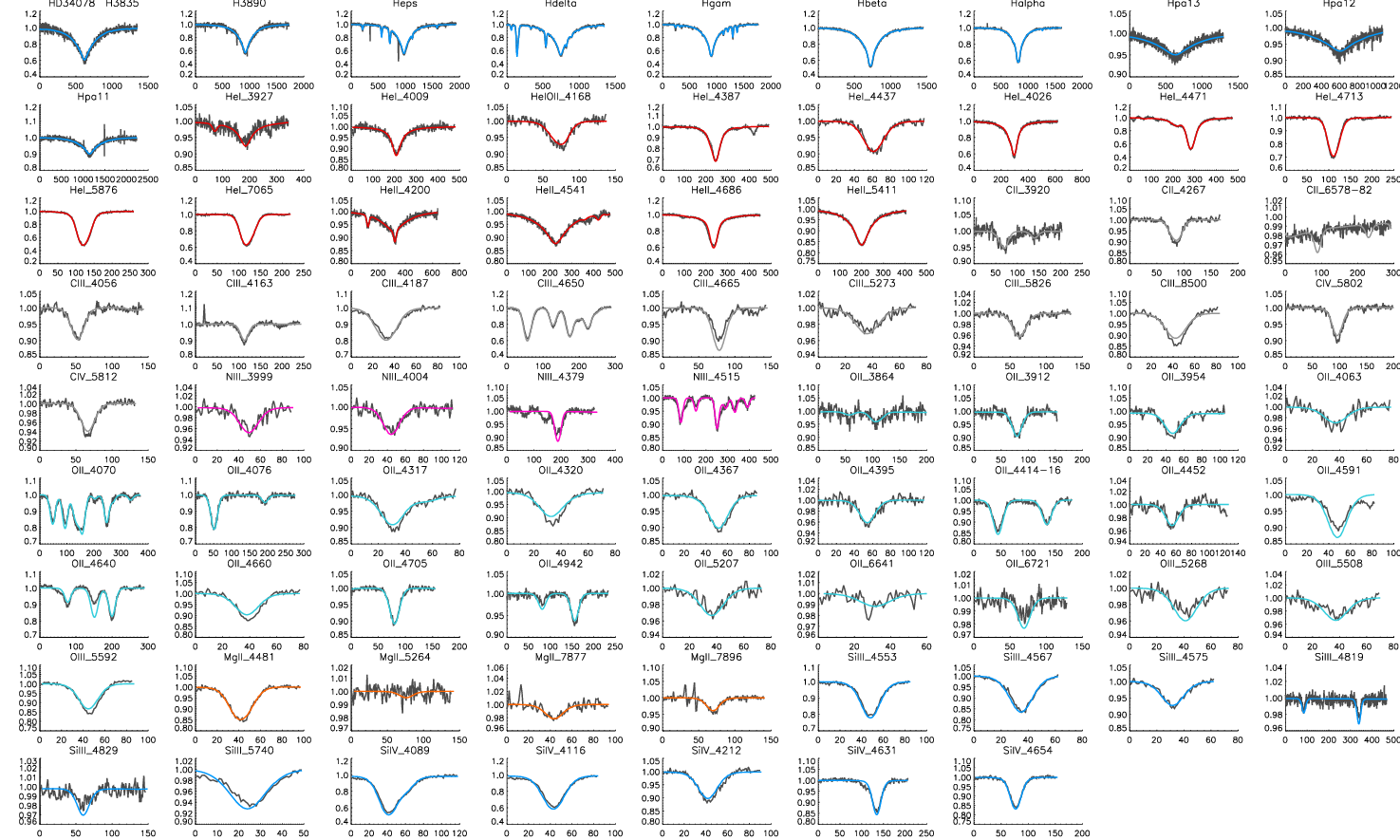}
\caption{HD\,34078 (O9.5\,V): comparison between the observed spectrum (black) and 
synthetic spectra (coloured) for selected diagnostic lines of H, He, C, N, O, 
Mg, and Si. The synthetic spectra were computed with \fastwind~using 
the stellar parameters and abundances obtained from the Bayesian analysis. 
The good agreement across multiple ions demonstrates the consistency of the 
inferred parameters and the robustness of the global fitting approach.
The horizontal axis is given in pixel number rather than wavelength; the corresponding spectral ranges are indicated by the panel labels. 
The names identifying the sub-panels are orientative only: each panel may include several spectral features, and the label marks approximately the central wavelength of the region shown.
}

  \label{fig:full_hd34078}
\end{figure*}

\begin{figure*}[ht]
 \centering
 \includegraphics[angle=90,width=0.7\textwidth]{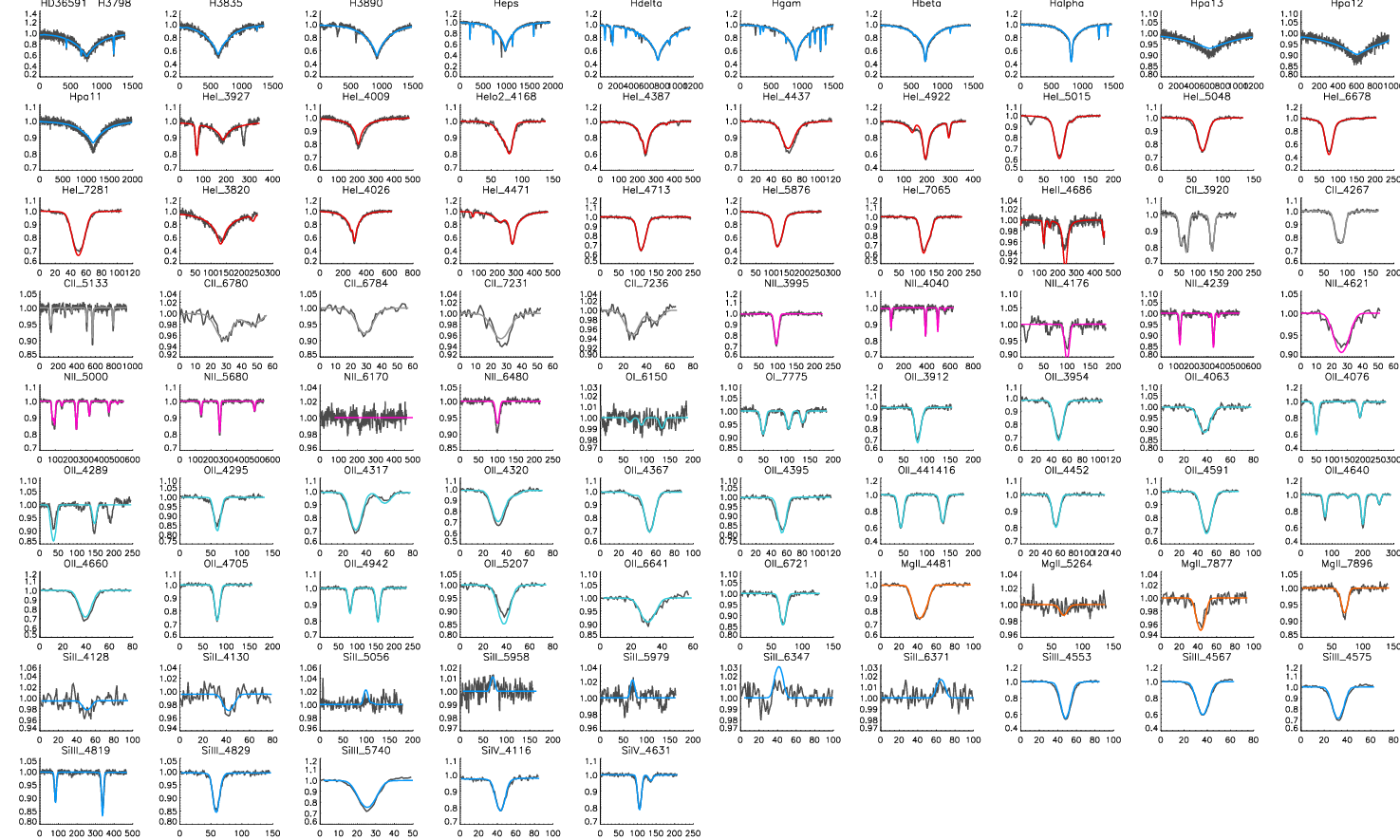}
\caption{As Fig.~\ref{fig:full_hd34078}, but for HD\,36591 (B1V).}
  \label{fig:full_hd36591}
\end{figure*}

The underlying \fastwind\ simulations include a much larger number of transitions, many of which are not explicitly included in the 
likelihood but are still predicted by the synthetic spectra. Some of these additional lines can be seen in the figures comparing the 
observed spectra with the model predictions (Figs.~\ref{fig:o_benchmark_fits}–\ref{fig:full_hd36591}), where they provide a qualitative consistency check on the overall model 
performance.

\begin{table}[ht!]
\caption{Spectral lines included in the likelihood evaluation. }
\centering
\begin{tabular}{ll}
\hline\hline
Element & Lines (\AA) \\
\hline
H       & Balmer series H$\epsilon$ to H$\alpha$  \\
He~{\sc i}  & 4026, 4387, 4471, 4713, 4922, 5015, 5048, 5876, 6678, 7065, 7281 \\
                  & (not used) 3820, 3868, 3889, 3927, 3965, 4009, 4024, 4120, 4143, 4437 \\  
He~{\sc ii}  & 4200, 4541, 5411, 6683 \\
C~{\sc ii}  & 3919, 3920, 4267, 6578, 6582 \\
C~{\sc iii}  & 4056, 4069, 4163, 4187, 4517, 4664, 4666, 4674, 5273, 5826, 8500 \\
C~{\sc iv}  & 5801, 5811 \\  
N~{\sc ii}  & 3995, 4004, 4035, 4041, 4447, 4607, 4803, 5001, 5005, 5011, 5026, 5045, 5676, 5680, 6480 \\
N~{\sc iii}  & 3999, 4003, 4511, 4515, 4518, 4523, 4527, 4634 \\
O~{\sc i}   & 6156-7-8, 7772-4-5 \\  
O~{\sc ii}  & 3792, 3913, 3955, 3963, 4277–78, 4284, 4305, 4318, 4321, 4368, 4416,  4418,  4592, 4597, \\ 
                &  4603, 4611, 4663, 4678, 4700, 4707, 4945, 6721 \\
O~{\sc iii}  & 4073, 4074, 4081, 4611, 5268, 5508, 7711 \\
Mg~{\sc ii}  & 4481, 7877, 7896 \\
Si~{\sc ii}  & 4128, 4130, 6347, 6371 \\
Si~{\sc iii}  & 4553, 4568, 4575, 4716, 4813, 4820, 4829, 5740 \\
Si~{\sc iv}  & 3762, 4116, 4212, 4631, 4654, 6668, 6701 \\
\hline
\end{tabular}
\label{app:lines}
\tablefoot{These features constitute the set of diagnostics used by \maui\ to constrain the stellar and wind parameters and chemical abundances.}
\end{table}

\FloatBarrier
\section{Detailed results}
This appendix presents the full set of tables and figures containing the results of the Bayesian analysis discussed in 
Sect.~\ref{sec:results}.

\begin{table}[ht!]
\centering
\caption{Fundamental parameters of the Galactic OB stars derived from our Bayesian analysis.}
\label{tab:fundamental_params}
\begin{tabular}{l l c c c c c c c}
\hline\hline
\# & Star & $T_{\mathrm{eff}}$ & $\log g$ & He/H & $\log Q$ & $\beta$ & $v\sin i$ & $v_{\mathrm{mac}}$ \\
   &           & (K)                         &  (dex)     &      &     (dex)    &        & (km\,s$^{-1}$) & (km\,s$^{-1}$) \\
\hline
1  & HD214680 & \unc{35400}{280}{200} & \unc{3.99}{0.05}{0.05} & \unc{0.10}{0.01}{0.01} & \unc{-14.87}{0.13}{0.44} & \unc{0.8}{0.1}{0.3} & \unc{14}{1}{1} & \unc{32}{1}{1} \\
2  & HD216898 & \unc{35700}{350}{430} & \unc{4.07}{0.07}{0.07} & \unc{0.09}{0.01}{0.01} & \unc{-14.49}{0.51}{0.32} & \unc{1.0}{0.3}{0.2} & \unc{44}{5}{5} & \unc{57}{6}{6} \\
3  & HD44597  & \unc{34400}{450}{450} & \unc{3.94}{0.08}{0.08} & \unc{0.12}{0.02}{0.02} & \unc{-14.26}{0.73}{0.35} & \unc{1.0}{0.3}{0.2} & \unc{14}{2}{2} & \unc{25}{6}{6} \\
4  & HD46202  & \unc{35000}{290}{220} & \unc{4.15}{0.05}{0.05} & \unc{0.09}{0.01}{0.01} & \unc{-14.93}{0.07}{0.19} & \unc{0.8}{0.1}{0.3} & \unc{11}{1}{1} & \unc{33}{4}{4} \\
5  & HD166546 & \unc{32600}{400}{480} & \unc{3.64}{0.08}{0.07} & \unc{0.09}{0.02}{0.01} & \unc{-13.56}{0.19}{0.49} & \unc{1.0}{0.3}{0.2} & \unc{33}{3}{3} & \unc{70}{6}{6} \\
6  & HD34078  & \unc{34500}{380}{360} & \unc{4.10}{0.08}{0.07} & \unc{0.12}{0.01}{0.01} & \unc{-13.50}{0.19}{0.34} & \unc{1.0}{0.3}{0.2} & \unc{9}{1}{1}  & \unc{23}{1}{1} \\
7  & HD161789 & \unc{33100}{380}{390} & \unc{4.07}{0.08}{0.07} & \unc{0.11}{0.01}{0.01} & \unc{-14.59}{0.41}{0.20} & \unc{1.0}{0.3}{0.2} & \unc{27}{3}{3} & \unc{15}{7}{7} \\
8  & HD36512  & \unc{33100}{300}{340} & \unc{4.12}{0.06}{0.07} & \unc{0.13}{0.01}{0.01} & \unc{-14.38}{0.59}{0.31} & \unc{1.0}{0.3}{0.2} & \unc{8}{4}{4}  & \unc{26}{4}{4} \\
9  & HD149438 & \unc{31900}{230}{250} & \unc{4.21}{0.06}{0.05} & \unc{0.12}{0.01}{0.01} & \unc{-14.62}{0.38}{0.15} & \unc{1.0}{0.3}{0.2} & \unc{4}{1}{1}  & \unc{4}{1}{1} \\
10 & HD36960  & \unc{29500}{380}{420} & \unc{4.11}{0.09}{0.08} & \unc{0.11}{0.01}{0.01} & \unc{-14.39}{0.58}{0.28} & \unc{1.0}{0.3}{0.2} & \unc{28}{2}{2} & \unc{20}{6}{6} \\
11 & HD36591  & \unc{27100}{360}{400} & \unc{4.11}{0.07}{0.06} & \unc{0.10}{0.01}{0.01} & \unc{-14.65}{0.34}{0.15} & \unc{1.0}{0.2}{0.1} & \unc{12}{1}{1} & \unc{0}{1}{1} \\
12 & HD36959  & \unc{25100}{460}{430} & \unc{4.24}{0.06}{0.06} & \unc{0.09}{0.01}{0.01} & \unc{-14.79}{0.21}{0.11} & \unc{0.9}{0.2}{0.1} & \unc{12}{1}{1} & \unc{5}{1}{1} \\
13 & HD35299  & \unc{23400}{240}{280} & \unc{4.24}{0.05}{0.05} & \unc{0.10}{0.01}{0.01} & \unc{-14.77}{0.23}{0.11} & \unc{0.9}{0.2}{0.1} & \unc{8}{1}{1}  & \unc{0}{1}{1} \\
14 & HD35039  & \unc{20200}{250}{260} & \unc{3.73}{0.05}{0.05} & \unc{0.10}{0.01}{0.01} & \unc{-14.77}{0.23}{0.13} & \unc{0.9}{0.2}{0.1} & \unc{12}{1}{1} & \unc{7}{1}{1} \\
15 & HD36430  & \unc{18600}{230}{230} & \unc{4.27}{0.04}{0.06} & \unc{0.10}{0.01}{0.01} & \unc{-14.03}{0.15}{0.26} & \unc{0.8}{0.1}{0.1} & \unc{20}{2}{2} & \unc{10}{2}{2} \\
16 & HD6675   & \unc{29300}{430}{350} & \unc{3.57}{0.08}{0.08} & \unc{0.09}{0.02}{0.01} & \unc{-14.31}{0.57}{0.36} & \unc{1.1}{0.1}{0.4} & \unc{24}{5}{5} & \unc{55}{4}{4} \\
17 & HD36822  & \unc{30300}{350}{390} & \unc{3.93}{0.08}{0.08} & \unc{0.10}{0.01}{0.01} & \unc{-14.23}{0.35}{0.51} & \unc{1.1}{0.1}{0.4} & \unc{28}{2}{2} & \unc{18}{5}{5} \\
18 & HD25443  & \unc{27700}{430}{330} & \unc{3.41}{0.08}{0.08} & \unc{0.10}{0.02}{0.02} & \unc{-14.09}{0.36}{0.53} & \unc{1.1}{0.1}{0.4} & \unc{32}{6}{6} & \unc{67}{4}{4} \\
19 & HD218376 & \unc{26600}{380}{360} & \unc{3.51}{0.08}{0.07} & \unc{0.12}{0.02}{0.02} & \unc{-14.28}{0.41}{0.51} & \unc{1.1}{0.1}{0.4} & \unc{25}{5}{5} & \unc{50}{4}{4} \\
20 & HD46328  & \unc{28500}{290}{290} & \unc{3.96}{0.07}{0.06} & \unc{0.09}{0.01}{0.01} & \unc{-14.59}{0.38}{0.18} & \unc{0.9}{0.2}{0.1} & \unc{5}{2}{2}  & \unc{12}{4}{4} \\
21 & HD201795 & \unc{29400}{240}{240} & \unc{4.28}{0.05}{0.07} & \unc{0.09}{0.01}{0.01} & \unc{-14.61}{0.36}{0.19} & \unc{1.0}{0.3}{0.2} & \unc{4}{4}{4}  & \unc{0}{0}{5} \\
22 & HD44743  & \unc{25100}{390}{410} & \unc{3.62}{0.08}{0.08} & \unc{0.11}{0.01}{0.01} & \unc{-14.33}{0.35}{0.51} & \unc{1.0}{0.3}{0.3} & \unc{16}{4}{4} & \unc{32}{5}{5} \\
23 & HD66665  & \unc{28300}{280}{260} & \unc{3.95}{0.05}{0.05} & \unc{0.11}{0.01}{0.01} & \unc{-14.73}{0.27}{0.15} & \unc{0.9}{0.2}{0.1} & \unc{7}{2}{2}  & \unc{4}{2}{2} \\
24 & HD16582  & \unc{22100}{260}{300} & \unc{3.91}{0.05}{0.07} & \unc{0.11}{0.01}{0.01} & \unc{-14.79}{0.21}{0.12} & \unc{1.0}{0.3}{0.2} & \unc{15}{2}{2} & \unc{10}{5}{5} \\
25 & HD886    & \unc{21700}{270}{220} & \unc{3.94}{0.05}{0.06} & \unc{0.09}{0.01}{0.01} & \unc{-14.74}{0.26}{0.15} & \unc{0.9}{0.2}{0.1} & \unc{9}{2}{2}  & \unc{8}{2}{2} \\
26 & HD35468  & \unc{21200}{480}{520} & \unc{3.59}{0.08}{0.09} & \unc{0.12}{0.01}{0.01} & \unc{-14.37}{0.59}{0.27} & \unc{1.0}{0.3}{0.2} & \unc{53}{10}{10} & \unc{27}{10}{10} \\
27 & HD160762 & \unc{18100}{200}{200} & \unc{3.98}{0.06}{0.06} & \unc{0.08}{0.01}{0.00} & \unc{-14.00}{0.22}{0.27} & \unc{1.0}{0.3}{0.2} & \unc{6}{1}{1}  & \unc{0}{0}{1} \\
\hline
\end{tabular}
\tablefoot{Quoted values are posterior medians with 68\% credible intervals.}
\end{table}

\begin{table}[ht!]
\centering
\caption{Chemical abundances of the Galactic OB stars derived from our Bayesian analysis.}
\label{tab:abundances}
\begin{tabular}{llcccccccc}
\hline\hline
\# & Star & $\xi$                & He/H  & C & N & O & Mg & Si \\
    &         & (km\,s$^{-1}$) &           & (dex) & (dex) & (dex) & (dex) & (dex) \\ 
\hline
1  & HD214680 & \unc{8}{1}{1} & \unc{0.10}{0.01}{0.01} & \unc{8.30}{0.05}{0.08} & \unc{7.95}{0.10}{0.07} & \unc{8.64}{0.06}{0.07} & \unc{7.49}{0.06}{0.05} & \unc{7.55}{0.10}{0.11} \\
2  & HD216898 & \unc{8}{2}{2} & \unc{0.09}{0.01}{0.01} & \unc{8.32}{0.11}{0.13} & \unc{7.75}{0.17}{0.16} & \unc{8.62}{0.12}{0.14} & \unc{7.45}{0.08}{0.09} & \unc{7.51}{0.20}{0.18} \\
3  & HD44597  & \unc{6}{1}{2} & \unc{0.12}{0.02}{0.02} & \unc{8.11}{0.10}{0.13} & \unc{7.82}{0.12}{0.14} & \unc{8.52}{0.10}{0.09} & \unc{7.46}{0.08}{0.10} & \unc{7.51}{0.22}{0.16} \\
4  & HD46202  & \unc{7}{1}{1} & \unc{0.09}{0.01}{0.01} & \unc{8.20}{0.07}{0.08} & \unc{7.92}{0.10}{0.09} & \unc{8.56}{0.08}{0.05} & \unc{7.45}{0.06}{0.05} & \unc{7.47}{0.12}{0.11} \\
5  & HD166546 & \unc{14}{2}{2} & \unc{0.09}{0.02}{0.01} & \unc{8.22}{0.12}{0.12} & \unc{7.64}{0.13}{0.18} & \unc{8.74}{0.07}{0.09} & \unc{7.52}{0.07}{0.08} & \unc{7.48}{0.12}{0.12} \\
6  & HD34078  & \unc{6}{1}{1} & \unc{0.12}{0.01}{0.01} & \unc{8.20}{0.07}{0.09} & \unc{7.70}{0.09}{0.09} & \unc{8.52}{0.07}{0.07} & \unc{7.52}{0.06}{0.05} & \unc{7.53}{0.13}{0.12} \\
7  & HD161789 & \unc{7}{2}{1} & \unc{0.11}{0.01}{0.01} & \unc{8.30}{0.10}{0.10} & \unc{7.80}{0.11}{0.12} & \unc{8.72}{0.07}{0.07} & \unc{7.56}{0.07}{0.07} & \unc{7.61}{0.16}{0.14} \\
8  & HD36512  & \unc{4}{1}{1} & \unc{0.13}{0.01}{0.01} & \unc{8.25}{0.08}{0.10} & \unc{7.80}{0.10}{0.10} & \unc{8.68}{0.07}{0.06} & \unc{7.48}{0.07}{0.06} & \unc{7.50}{0.17}{0.12} \\
9  & HD149438 & \unc{3}{1}{1} & \unc{0.12}{0.01}{0.01} & \unc{8.21}{0.08}{0.08} & \unc{8.18}{0.06}{0.07} & \unc{8.62}{0.04}{0.04} & \unc{7.47}{0.05}{0.05} & \unc{7.56}{0.06}{0.07} \\
10 & HD36960  & \unc{5}{1}{1} & \unc{0.11}{0.01}{0.01} & \unc{8.14}{0.11}{0.12} & \unc{7.62}{0.12}{0.10} & \unc{8.64}{0.06}{0.06} & \unc{7.45}{0.07}{0.07} & \unc{7.45}{0.12}{0.12} \\
11 & HD36591  & \unc{3}{1}{1} & \unc{0.10}{0.01}{0.01} & \unc{8.20}{0.08}{0.10} & \unc{7.74}{0.09}{0.07} & \unc{8.78}{0.04}{0.05} & \unc{7.43}{0.06}{0.06} & \unc{7.44}{0.10}{0.10} \\
12 & HD36959  & \unc{1}{1}{1} & \unc{0.09}{0.01}{0.01} & \unc{8.17}{0.09}{0.12} & \unc{7.81}{0.08}{0.10} & \unc{8.83}{0.08}{0.09} & \unc{7.41}{0.06}{0.06} & \unc{7.54}{0.10}{0.12} \\
13 & HD35299  & \unc{1}{1}{1} & \unc{0.10}{0.01}{0.01} & \unc{8.22}{0.06}{0.07} & \unc{7.82}{0.06}{0.06} & \unc{8.86}{0.06}{0.08} & \unc{7.44}{0.05}{0.04} & \unc{7.49}{0.10}{0.08} \\
14 & HD35039  & \unc{1}{1}{1} & \unc{0.10}{0.01}{0.01} & \unc{8.34}{0.06}{0.07} & \unc{7.76}{0.09}{0.07} & \unc{8.85}{0.07}{0.09} & \unc{7.58}{0.06}{0.05} & \unc{7.55}{0.09}{0.09} \\
15 & HD36430  & \unc{0}{0}{1} & \unc{0.10}{0.01}{0.01} & \unc{8.41}{0.09}{0.10} & \unc{7.88}{0.15}{0.18} & \unc{8.65}{0.22}{0.28} & \unc{7.41}{0.04}{0.05} & \unc{7.39}{0.12}{0.11} \\
16 & HD6675   & \unc{14}{1}{1} & \unc{0.09}{0.02}{0.01} & \unc{8.07}{0.11}{0.11} & \unc{7.62}{0.10}{0.11} & \unc{8.63}{0.05}{0.05} & \unc{7.46}{0.05}{0.07} & \unc{7.47}{0.09}{0.07} \\
17 & HD36822  & \unc{8}{1}{1} & \unc{0.10}{0.01}{0.01} & \unc{8.12}{0.10}{0.10} & \unc{7.86}{0.09}{0.09} & \unc{8.63}{0.06}{0.05} & \unc{7.49}{0.07}{0.05} & \unc{7.57}{0.11}{0.09} \\
18 & HD25443  & \unc{15}{2}{1} & \unc{0.10}{0.02}{0.02} & \unc{8.11}{0.12}{0.13} & \unc{7.56}{0.10}{0.12} & \unc{8.61}{0.05}{0.06} & \unc{7.49}{0.06}{0.08} & \unc{7.48}{0.10}{0.09} \\
19 & HD218376 & \unc{12}{1}{1} & \unc{0.12}{0.02}{0.02} & \unc{8.15}{0.11}{0.10} & \unc{7.97}{0.08}{0.07} & \unc{8.66}{0.05}{0.04} & \unc{7.47}{0.06}{0.06} & \unc{7.48}{0.10}{0.09} \\
20 & HD46328  & \unc{3}{1}{1} & \unc{0.09}{0.01}{0.01} & \unc{8.12}{0.09}{0.09} & \unc{8.11}{0.06}{0.06} & \unc{8.78}{0.04}{0.04} & \unc{7.58}{0.06}{0.05} & \unc{7.63}{0.09}{0.07} \\
21 & HD201795 & \unc{2}{1}{1} & \unc{0.09}{0.01}{0.01} & \unc{8.28}{0.09}{0.09} & \unc{7.74}{0.06}{0.07} & \unc{8.74}{0.04}{0.03} & \unc{7.45}{0.05}{0.06} & \unc{7.58}{0.07}{0.06} \\
22 & HD44743  & \unc{10}{1}{1} & \unc{0.11}{0.01}{0.01} & \unc{8.17}{0.09}{0.10} & \unc{7.63}{0.07}{0.08} & \unc{8.69}{0.05}{0.05} & \unc{7.42}{0.05}{0.06} & \unc{7.46}{0.10}{0.09} \\
23 & HD66665  & \unc{1}{1}{1} & \unc{0.11}{0.01}{0.01} & \unc{8.01}{0.06}{0.07} & \unc{7.99}{0.06}{0.05} & \unc{8.79}{0.03}{0.04} & \unc{7.44}{0.04}{0.05} & \unc{7.52}{0.07}{0.07} \\
24 & HD16582  & \unc{1}{1}{1} & \unc{0.11}{0.01}{0.01} & \unc{8.17}{0.09}{0.09} & \unc{8.23}{0.09}{0.09} & \unc{8.79}{0.08}{0.06} & \unc{7.54}{0.05}{0.05} & \unc{7.63}{0.10}{0.09} \\
25 & HD886    & \unc{0}{0}{1} & \unc{0.09}{0.01}{0.01} & \unc{8.30}{0.09}{0.09} & \unc{7.82}{0.09}{0.09} & \unc{8.81}{0.07}{0.07} & \unc{7.56}{0.05}{0.05} & \unc{7.61}{0.08}{0.10} \\
26 & HD35468  & \unc{5}{1}{1} & \unc{0.12}{0.01}{0.01} & \unc{8.17}{0.14}{0.13} & \unc{8.10}{0.16}{0.13} & \unc{8.75}{0.11}{0.12} & \unc{7.49}{0.10}{0.08} & \unc{7.58}{0.19}{0.18} \\
27 & HD160762 & \unc{2}{1}{1} & \unc{0.08}{0.01}{0.01} & \unc{8.25}{0.09}{0.09} & \unc{7.66}{0.15}{0.21} & \unc{8.66}{0.10}{0.10} & \unc{7.42}{0.06}{0.05} & \unc{7.42}{0.08}{0.10} \\
\hline
\end{tabular}
\tablefoot{Quoted values are posterior medians with 68\% credible intervals. Abundances are given as 
$\log \epsilon(X) = \log (N_X/N_H) + 12$.}
\end{table}

\end{appendix}

\end{document}